# *H*-linear magnetoresistance in $NbSe_2$ due to impeded cyclotron motion

Arwin Kool,[1,2]* Davide Pizzirani,[1,2] Paul Tinnemans,[2] Steffen Wiedmann,[1,2] Felix Flicker,[3] Jasper van Wezel,[4] Nigel E. Hussey,[1,2,3]* Roemer D. H. Hinlopen[3]*

**Affiliations**

[1]High Field Magnet Laboratory (HFML-FELIX), Radboud University, Nijmegen, Netherlands
[2]Institute for Molecules and Materials, Radboud University, Nijmegen, Netherlands
[3]H. H. Wills Physics Laboratory, University of Bristol, Bristol, United Kingdom
[4]Institute for Theoretical Physics, University of Amsterdam, Amsterdam, Netherlands

*To whom correspondence should be addressed. E-mail: arwin.kool@ru.nl, n.e.hussey@bristol.ac.uk or roemer.hinlopen@mpsd.mpg.de.

**Abstract**

Linear magnetoresistance (LMR) is a widespread phenomenon observed in a host of quantum materials ranging from semiconductor nanostructures to quantum critical and strange metals. While multiple scenarios to explain LMR have been proposed, a complete understanding of the phenomenon remains elusive. Indeed, it is highly likely that the origin of LMR depends on the specific electronic state. Here, we report a study of the impact of disorder on the form of the magnetoresistance of the prototypical charge-density-wave (CDW) compound 2*H*-$NbSe_2$. The magnetoresistance is shown to exhibit strong qualitative and quantitative agreement with Boltzmann transport analysis incorporating impeded cyclotron motion (ICM). We identify the source of ICM in 2*H*-$NbSe_2$ as strong scattering sinks where the CDW order connects the high-temperature Fermi cylinders. Such unusual 'hotspots' provide an explanation for the observed LMR as well as for the long-unexplained absence of quantum oscillations inside the charge-ordered state in 2*H*-$NbSe_2$. These findings provide strong evidence that ICM generates LMR in certain correlated metals.

**Teaser**

Linear magnetoresistance in $NbSe_2$ is shown to be linked to impeded cyclotron motion induced by the low-temperature charge order.

**Introduction**

Linear-in-field magnetoresistance (LMR) contrasts markedly with the usual quadratic-to-saturating shape of the MR found in conventional metals. This, coupled with its remarkable robustness (extending over an anomalously broad field range) and its realization in a host of different quantum materials, from semiconductors (*1*,*2*) and Dirac semi-metals (*3*) to quantum critical metals (*4*-*6*) and doped Mott insulators (*7*-*9*), has led to sustained interest in the LMR phenomenon for well over half a century. Reflecting the breadth of material classes in which LMR has been observed, multiple proposals have been put forward for its origin. The first category of proposals centers on materials with a low or vanishing carrier density. In GaAs quantum wells, for example, carrier density fluctuations result in a non-saturating LMR up to high magnetic fields (*1*,*2*). For more strongly disordered semiconductors, LMR can arise from the formation of an effective random resistor network (*10*,*11*), while LMR is also expected at high magnetic field strengths near the quantum limit, where only a few Landau levels remain occupied (*12*). For metals with a high carrier density, LMR can arise, albeit over a limited field range, from sharp corners on the Fermi surface (FS) (*13*-*15*).

The manifestation of LMR in correlated metals with relatively simple, large, cylindrical FS geometries, however, suggests an alternative origin possibly linked to the emergence of novel electronic states. In iron-based superconductors tuned to their quantum critical point, for example, a peculiar form of non-saturating $H^2$- to $H$-linear MR - with a $T$-independent slope - has been reported (*4-6*). The MR in these systems also exhibits a novel form of (non-Kohler) scaling in which magnetic field and temperature (rather than the zero-field resistivity) appear in quadrature (*4*). LMR with similar scaling properties has also been observed in the high-$T_c$ cuprates, both close to the end of the pseudogap regime (*7*) and beyond (*8*,*9*). In the latter, the $T$-independent slope of the LMR is found to scale directly with both $T_c$ and the coefficient of the low-$T$ $T$-linear resistivity (*16*), suggesting a direct link to strange metallicity and high-$T_c$ superconductivity.

The simplicity and robustness of the LMR found in the iron-based and cuprate superconductors has motivated a number of tailored theoretical proposals, including real-space binary distributions (*17*), real-space patches (*18*), carrier density fluctuations (*19*) and strong scattering sinks (*20-22*) among others. Proposals that incorporate scattering 'hotspots' or any other obstruction to cyclotron motion on the FS can be generalized through the concept of impeded cyclotron motion (ICM) (*21*). In this picture, the orbital motion of carriers around the FS is truncated at specific loci, for example through gapping, heavy effective masses or intense scattering. These loci themselves are irrelevant for the resistivity of the material in the absence of a magnetic field due to shorting by the rest of the FS (which we will refer to as cold). With field applied, however, cyclotron motion of the cold carriers is impeded, resulting in a characteristic crossover from $H^2$- to $H$-linear MR with a slope that is $T$-independent (*21*). The universality of LMR in correlated metals is then explained by the wide range of possible sources of the impedances that might occur, including hotspot scattering off antiferromagnetic spin fluctuations (*20*,*23*), van Hove singularities (*24*), static domains of glassy density-wave order (*22*), Fermi arcs in underdoped cuprates (*25*) or partial FS incoherence (*9*). Hence, identifying materials where ICM can account for the form and magnitude of the LMR as well as the range in tuning parameters where LMR is observed would provide definitive evidence of a dominant (and strongly $k$-dependent) interaction.

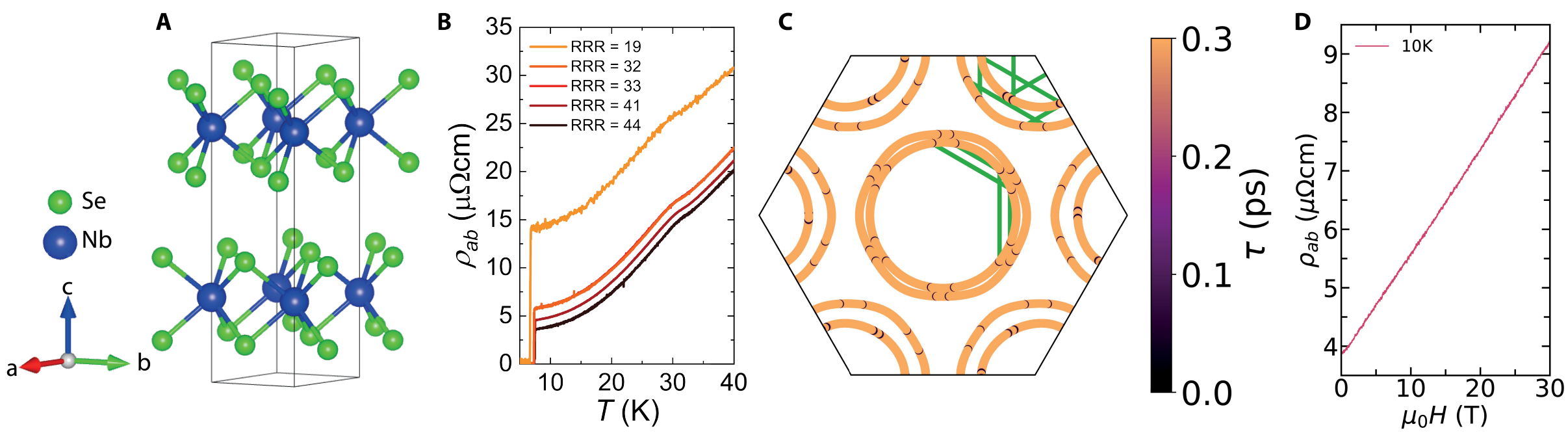

**Fig. 1. Lattice, resistivity versus temperature and simulation of the FS of bulk $NbSe_2$ in the charge density wave (CDW) state. A)** Crystal structure of bulk $2H$-$NbSe_2$. The crystal structure consists of two weakly coupled, hexagonal Se-Nb-Se layers. **B)** In-plane resistivity $\rho_{ab}(T)$ for five samples of $NbSe_2$ of varying quality as indicated by the factor of 4 variation in the residual resistivity $\rho_0$. The superconducting and CDW transition temperatures decrease monotonically with increasing $\rho_0$ (decreasing residual resistivity ratio or *RRR*, consistent with previous disorder studies (37,38). **C)** The tight-binding FS of bulk $NbSe_2$ previously determined via ARPES (*35*). Unless mentioned otherwise, we ignore the small pancake-shaped pocket centered around the Γ-point. Instead of a FS reconstruction, impedances (black) are shown where the FS is connected intrapocket by the CDW $\vec{Q}$-vector (examples in green) while cold charge is colored orange. **D)** Exemplary LMR measured on the sample with $RRR$ = 44 at $T$ = 10 K.

In this context, the prototypical charge-density-wave (CDW) compound 2*H*-$NbSe_2$ (henceforth abbreviated to $NbSe_2$) can be considered a viable candidate for the realization of ICM (*21*). Among the dichalcogenides, $NbSe_2$ has a relatively low CDW transition temperature of $T_{CDW}$ = 33 K and a high superconducting (SC) transition temperature of $T_c$ = 7.2 K. As shown in Fig. 1**A**, $NbSe_2$ is comprised of Nb slabs sandwiched between Se atoms. LMR was first reported in $NbSe_2$ more than 50 years ago (*26*) and ascribed to magnetic breakdown through the gaps opened by the CDW (*27*). Such behavior, however, is known to result in MR saturation rather than LMR (*28*-*30*). The lack of quantum oscillations (QOs) from all pockets which participate in the CDW order also remains striking and unexplained (*31*,*32*). However, if the CDW order manifests with a large imaginary self-energy for any reason (e.g. low single-particle lifetimes for phonon-dressed quasiparticles), then the resulting impedances to cyclotron motion could explain both a lack of QOs on the participating pockets and the emergence of LMR which disappears with increasing temperature or pressure in tandem with the CDW phase (*33*).

In order to test the applicability of ICM to $NbSe_2$, we have carried out a magnetotransport study of bulk $NbSe_2$ over an unprecedented field range as a function of temperature and disorder. The LMR is found to disappear upon approach to the CDW transition temperature from below and, to within our experimental accuracy, follows Kohler scaling up to this breakdown temperature. Recall that Kohler scaling is a single-parameter scaling of the magnetoresistance with $\omega_c\tau$ (*34*) that is expected semiclassically if the MR changes solely due to a *T*-dependent scattering rate. Kohler scaling is obeyed in most elemental metals, but can be violated, for instance, through changes in anisotropy or in the size of the Fermi surface, by a field dependence of the anisotropy or of the scattering rate, through changes in the effective mass and by changes in the mobility ratio between different pockets. Furthermore, no quantum oscillations have been observed in the cleanest crystals up to the highest field strength (30 T) and down to lowest temperatures (0.35 K). Using the known FS of $NbSe_2$ (*35*) and a single degree of freedom (the zero-field scattering rate determined by impurity scattering), we are able to explain the absence of QOs and model the presence of the LMR within the ICM framework. The slope of the *H*-linear MR agrees quantitatively with the model's prediction, while Kohler scaling and the temperature and disorder independence of the LMR slope are also reproduced. For completeness, limitations of the model and effects beyond the relaxation time approximation are also discussed.

## Results

### Disorder dependence of the high-field MR of $NbSe_2$

Fig. 1**B** shows the low-*T* $\rho(T)$ curves for five single crystals of $NbSe_2$ with a 4-fold variation in their residual resistivity $\rho_0 = \rho(T_c)$. As reported in previous disorder studies, both $T_c$ and $T_{CDW}$ are found to drop with increasing $\rho_0$ (*36*-*39*). The CDW opens gaps primarily where parts of the FS are connected by the CDW wave vector $\vec{Q}$ within an individual pocket, and most strongly on the inner *K*- and *K*'-barrels (*40*-*42*). The ICM hypothesis applied to $NbSe_2$ is that where such gaps open, additional sharp drops in the transport lifetime $\tau$ are experienced by the quasiparticles (Fig. 1**C**). This drop in lifetime is assumed to scale with the strength of the CDW order, disappearing at the CDW transition. An exemplar LMR within the CDW phase is shown in Fig. 1**D**. Below, we use the FS previously determined via ARPES (*35*) and this scattering lifetime to model the MR of $NbSe_2$ using a Boltzmann transport calculation. Additionally, in order to investigate specific limitations of the model, we present calculations that include the proposed FS reconstruction as well as calculations that go beyond the relaxation time approximation.

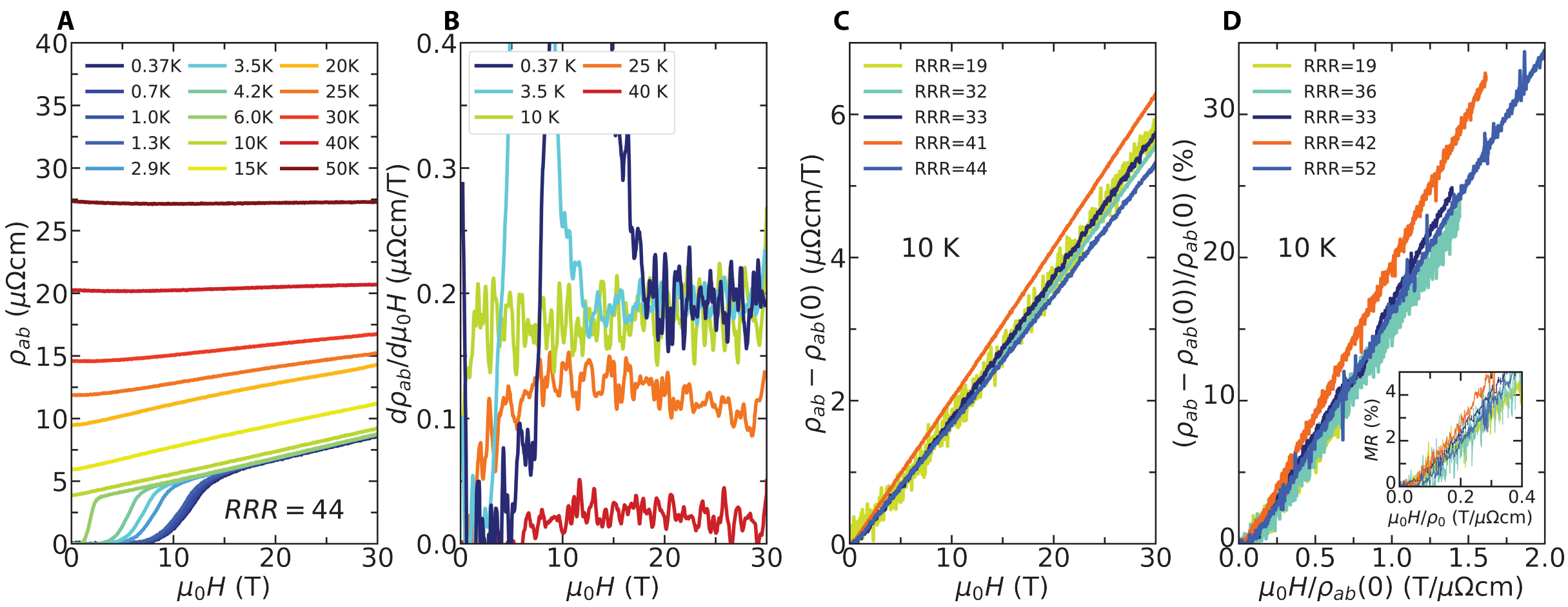

**Fig. 2. Overview of the magnetoresistance behavior in 2$H$-NbSe$_2$. A)** Raw MR data for the sample with the largest *RRR* measured between 0.37 and 50 K. The current is applied along the *a*-axis, and the magnetic field is oriented in the *c*-direction. We observe a clear *H*-linear MR over an extended field range below 15 K. Above 15 K, the magnitude decreases until the MR effectively vanishes at 50 K. **B)** Field derivative of the resistivity $d\rho_{ab}/d\mu_0H$ reveals a robust non-saturating, *T*-independent LMR below *T* =15 K that persists up to the highest magnetic fields. Above 15 K, the MR develops a tendency towards saturation. **C)** MR curves obtained just above $T_c$ (at 10 K) in the crystals whose zero-field resistivity is shown in Fig. 1B). The high field MR is clearly linear, while the low field MR shows a more quadratic character. The field range of the quadratic behavior depends on the *RRR*. **D)** Kohler plot for field sweeps made up to 8 T of crystals cut from the same mother crystals as the high-field samples, but with slightly different *RRR*'s. Kohler scaling is obeyed at 10 K as a function of disorder. Inset: zoom of the low-field quadratic part of the MR, showing the robustness of the scaling.

Fig. 2**A** shows a representative set of $\rho_{ab}$ measurements as a function of applied magnetic field for the sample with *RRR* = 44, where *RRR* = $\rho_{ab}$(RT)/$\rho_{ab}(T_c)$. At the lowest temperatures and beyond a magnetic field strength large enough to suppress superconductivity, we observe *H*-linearity up to 30 T. Between $T_c$ and 15 K, we observe a $H^2$- to *H*-linear crossover in the MR with a small crossover scale $H^*$ between 0.5 and 1.0 T. Above 15 K, the *H*-linearity gradually breaks down, then above $T_{CDW}$, the MR almost vanishes and exhibits more conventional quadratic-to-saturating behavior. Correspondingly, the Hall effect (see Fig. S1) is temperature independent and electron-like below 15 K, changes sign at 25 K and recovers the expected hole-like linear Hall response above ~ 40 K. At low-*T*, the *H*-linear slope is independent of temperature and has a value ≈ 0.18 $\mu\Omega$cm/T (see Fig. 2**B**). This compares with previous literature values of 0.10 (*43*) and 0.13 $\mu\Omega$cm/T (*27*).

Fig. 2**C** shows $\Delta\rho_{ab}$ (= $\rho_{ab}(\mu_0H)$-$\rho_{ab}(0)$) of the five measured samples at *T* = 10 K. (The full dataset for all samples can be found in Fig. S1). The *H*-linear slope of the MR is found to vary by ±3% despite a 4-fold variation in $\rho_0$. To properly test Kohler scaling, the $H^2$-part of the MR should also collapse, which is obscured in Fig. 2**C** due to the low field scale of the turnover. We therefore performed a low-field study on crystals cut from the same mother crystals as the high field samples, but with slightly different *RRR*'s, to test specifically whether the quadratic part of the MR curves follows any type of scaling. The full dataset of these measurements is presented in Fig. S2 and Fig. S3. We show the Kohler plot of these samples measured at *T* = 10 K in Fig. 2**D**. In the inset of Fig. 2**D**, we also show a blow-up of the same Kohler plot, focusing on the low-field response. As can be seen, the curves collapse to within experimental error, implying that the MR in $NbSe_2$ does indeed obey Kohler scaling. In the following section, we proceed to model the data to better understand the origin of the LMR in $NbSe_2$.

**Modelling the data using impeded cyclotron motion**

ICM provides a phenomenological description of MR that is insensitive to the microscopic details. This allows us to thoroughly test the model against the experimental data with few degrees of freedom. We follow Ref. (*21*) and hypothesize that the predominant effect of the CDW order on the electron dynamics is to turn points on the FS connected by a CDW $\vec{Q}$-vector incoherent or, equivalently, to enlarge their imaginary self-energies into a set of hotspots. The suppression of the carrier lifetime at the hotspots then scales with the strength of the CDW order and impedes cyclotron motion, suppressing QOs and generating LMR.

Even at $\mu_0 H$ = 30 T and $T$ = 0.35 K, QOs associated with the major pockets (with predicted frequencies of 5-10 kT) are absent. Under these conditions, they would be expected to trace out a real-space orbit of typical cyclotron radius $R_c \approx$ 75 nm based on the known FS of $NbSe_2$. This is in stark contrast with the pancake pocket, which has been observed at 2.5 T when $R_c$ = 225 nm despite being subject to additional scattering in the vortex liquid state (*31*,*32*). We similarly observe QOs from the pancake pocket when we rotate the magnetic field, see Fig. S5. The cyclotron radius of the smallest pockets of $NbSe_2$ after backfolding and hybridization is expected to be as small as $R_c \approx$ 8 nm in our experimental conditions (see the SM for details) yet no quantum oscillations are observed in these either. This striking lack of quantum oscillations supports the notion that the CDW order in 2*H*-$NbSe_2$ forms impedances at specific positions in *k*-space, preventing electrons from completing cyclotron orbits anywhere on the FS (except around the pancake pocket that of itself does not participate in the CDW order).

In order to test this hypothesis more rigorously, we develop a phenomenological magneto-transport model using the minimal tight-binding expansion of the FS of $NbSe_2$ (*35*). Backfolding and hybridization undoubtedly occur, but by themselves do not explain the observations and so we neglect these steps initially to focus on the essential ingredients required to explain both the MR and the absence of quantum oscillations. Impedances are placed at locations at which the CDW $\vec{Q}$-vector connects the FS, only keeping intra-pocket connections in accordance with Ref. (*40*) and noting that the spectral weight observed at other locations on the FS in ARPES remains unchanged (*35*,*44*). The impedances are modeled as locally suppressed lifetimes as shown in Fig. 1**C**. We use the lifetime for its computational simplicity, but emphasize that the results are valid also if the impedances result, e.g. from a removal of spectral weight away from the Fermi energy or from high local effective masses; in all cases the contribution of the impedances themselves to the total conductivity is negligible (due to shorting effects) and affects the magnetotransport solely by ICM. We use the full Shockley-Chambers Tube Integral Formalism (SCTIF) of the Boltzmann transport equation assuming the relaxation time approximation and neglecting *c*-axis dispersion (*45*,*46*). The only degree of freedom in the model is the isotropic cold scattering time $\tau$ across the FS, which we fit to the zero-field resistivity $\rho_{ab}(0)$.

We start with the most general solution to the Boltzmann transport equation within the relaxation time approximation and without thermal broadening. The resulting formula within the SCTIF is (*45*,*46*):

$$\sigma_{ij} = \frac{e^2}{\pi\hbar} \int_{\mathrm{FS}} \frac{\mathrm{d}^2\vec{k}}{(2\pi)^2} \frac{v_i}{v} \int_0^\infty \mathrm{d}t v_j(-t) \exp\left( - \int_0^t \frac{\mathrm{d}t'}{\tau\left(-t'\right)} \right) \tag{1}$$

Here, $\sigma$ is the conductivity, indices *i*,*j* run over *x*,*y* (the $k_z$ corrugation of the Fermi surface is neglected), *e* is the electron charge, $\hbar$ the reduced Planck's constant, FS the Fermi surface including a sum over the different pockets, $\tau$ the velocity relaxation time and $\vec{v} := \frac{1}{\hbar}\vec{\nabla}_k \varepsilon_k$ is the Fermi velocity defined through the energy dispersion $\varepsilon_k$. The $\vec{k}$ dependence of $\vec{v}$ and $\tau$ is implicit for clarity. Time dependence manifests through cyclotron motion, i.e. through $\vec{k}(t)$. The exponent in Eq. (1) is the probability for a quasiparticle to survive to time $t$. The integrals can be evaluated for arbitrary magnetic field strengths. Temperature dependence arises solely from the $T$-dependence of $\tau$. Resistivity is obtained through a full matrix inversion of $\sigma$. The model is defined entirely by $\varepsilon_k$ and $\tau(\vec{k})$. We take $\varepsilon_k$ from a previous tight-binding fit to ARPES data which contains exactly 1 hole per unit cell while neglecting the selenium pancake pocket (*35*). $\tau(k)$ is shown in Fig. 1C for the entire Fermi surface and in Fig. 3D for an individual hotspot.

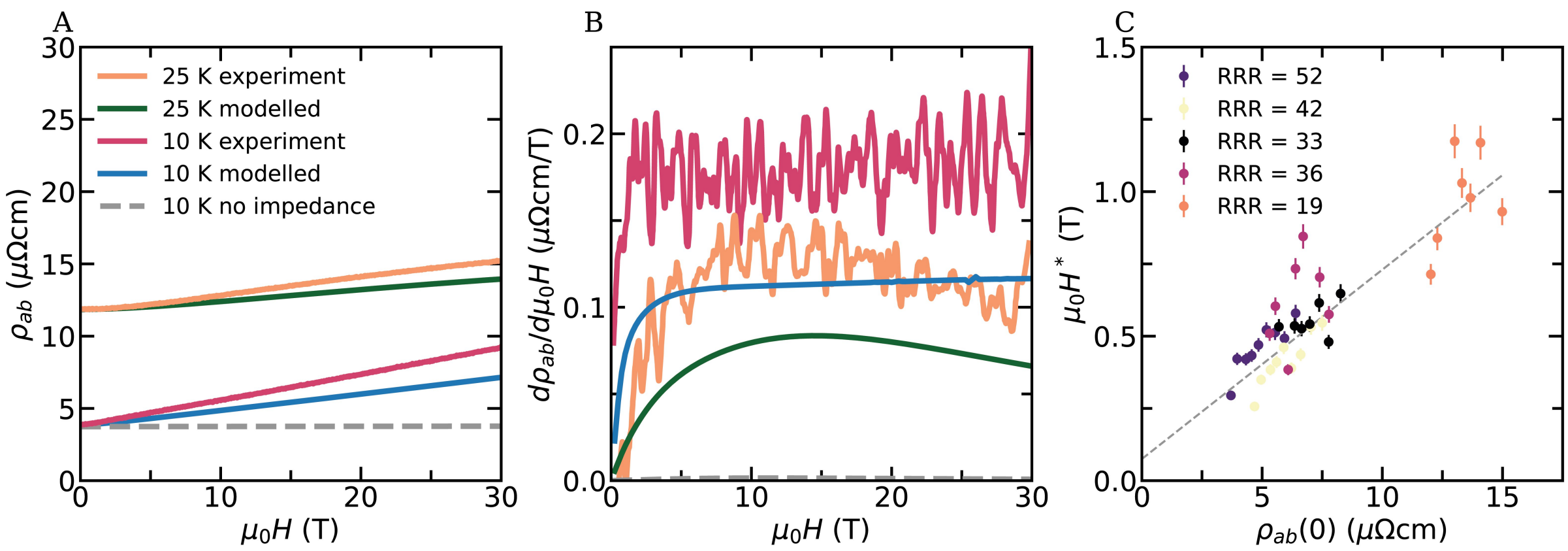

**Fig. 3. Comparison of the ICM model to the measured data. A)** Comparison of the modelled MR response (blue) to the measured data (purple) on the *RRR* = 44 sample at $T$ = 10 K, i.e. well within the $H$-linear regime. The modelled response without impedances (gray) indicates that the MR originates from the impedances and not from the FS itself. A second comparison (green and orange, respectively) is shown at $T$ = 25 K (close to $T_{\mathrm{CDW}}$) where we observe a breakdown of $H$-linearity. This breakdown at high magnetic field can be reproduced by artificially weakening the hotspots such that quasiparticles have a finite probability for breakdown, see SI for details. **B)** The derivatives of the curves shown in panel A) highlight the linearity and turnover scale. Overall, the model shows good qualitative and reasonable quantitative agreement with the data. **C)** $H^*$ versus $\rho_{\mathrm{ab}}(0)$ obtained by varying both disorder and temperature between 8 and 18 K. Here, $H^*$ is extracted as the field value where the derivative $d\rho_{\mathrm{ab}}$ $/d\mu_0 H$ reaches 90% of the high-field linear slope. The dashed gray line is a guide to the eye. Representative impedances at 10 K and 25 K. $\phi$ refers to the azimuthal angle on the inner $\Gamma$ pocket. By weakening impedances with increasing temperature towards $T_{\mathrm{CDW}}$, the observed MR saturation is captured.

In panels **A** and **B** of Fig. 3 we directly compare the simulations obtained from our simple model to the experimental data obtained on the sample with *RRR* = 44. For clarity, we only include curves at $T$ =10 K and 25 K. As shown in Fig. 3**B** at $T$ =10 K, the model generates a robust LMR with a slope of 0.11 $\mu\Omega$cm/T that persists up to the highest fields used in our study. Consistent with the data shown in Fig. 2**C**, this $H$-linear slope is independent of $\tau$ ($\rho_{ab}(0)$). Despite the lack of tunability, the model captures well the crossover field $H^*$ and reproduces the magnitude of the MR to within 30%.

The LMR slope in ICM is nominally inversely proportional to the cold carrier density (akin to the traditional Hall effect) and robust against changes in scattering time, temperature, disorder or effective mass as observed experimentally. Upon increasing temperature, the CDW gap decreases and as a result, the impedances are expected to disappear. Consistent with ICM, the LMR is suppressed near $T_{\mathrm{CDW}}$. In fact, the experimentally observed MR essentially vanishes at 50 K and the observed and modeled Hall effect match with expectations from the carrier density (see the SM for details). In the intermediate temperature range just below $T_{\mathrm{CDW}}$, we anticipate the CDW order to be sufficiently weak that carriers are affected by the CDW, but their cyclotron motion may not be effectively impeded. (Note that given the estimated effective mass in $NbSe_2$, this temperature is too high for QOs to be observed.) To incorporate this effect, we introduce a second degree of freedom by artificially weakening the impedances such that magnetic breakdown is possible within the experimentally observable regime. This means that the hotspots are not infinitely strong and there remains a finite probability exp(-$B_{br}/B$) for charge to maintain cyclotron motion across the impedance. At low $T$, the breakdown field $B_{br}$ is much greater than 50 T and decreases towards $T_{CDW}$. The result for $T$ = 25 K is shown in Fig. 3**A** and Fig. 3**B** and reproduces the observations: The MR diminishes in size and saturates in tandem. In other words, the robust LMR has vanished. Note that the tendency of the MR in $NbSe_2$ to saturate already at intermediate temperatures is unusual yet is well reproduced by the ICM model. Recent results at low $T$ as a function of hydrostatic pressure show that the MR similarly diminishes together with the CDW order (*33*). Collectively, these experimental observations appear to tie the LMR to the CDW order, which is naturally explained by impedances which scale with the CDW order parameter as adopted here. Finally, the model naturally reproduces Kohler scaling at low $T$ where the isotropic cold scattering rate is the only temperature dependent parameter.

The turnover scale $H^*$ where LMR sets in represents the magnetic field at which cold quasiparticles can traverse between hotspots via cyclotron motion. For this process, we expect the Kohler relation $H^* \propto \rho(0)$ to hold both as a function of temperature (below ~ ⅔$T_{CDW}$ when impedances appear to be well established and breakdown seems absent) as well as disorder. This relationship, shown in Fig. 3**C**, agrees well with the model predictions. In the SM, we present a version of the model that also fits the result in absolute units for all temperatures, magnetic fields and disorder values by incorporating not only the cold scattering rate, but also the width and strength of hotspots to account for magnetic breakdown.

Despite its successes, this simple model has some shortfalls that we summarize here while referring the reader to the SM for more details. First and foremost, although ICM describes a simultaneous $H$-linearity in the MR and Hall effect, the sign change in the Hall effect which sets in simultaneously with the LMR remains unexplained. Secondly, the model relies on the relaxation time approximation and thus neglects contributions to the conductivity from non-zero averaged velocities near the hotspots (*47*). To address these points, we consider in the SM various extensions to the model that include the full FS reconstruction, additional anisotropy in the reconstructed phase from interpocket and interband interactions and a practical Boltzmann transport framework beyond the relaxation time approximation that was previously applied to quasi-1D conductors in Ref. (*48*). Firstly, we find that the Peierls reconstruction without impedances is insufficient to explain the absence of quantum oscillations or the presence of LMR in $NbSe_2$. Secondly, we address any concerns about charge conservation within impeded cyclotron motion by developing a fully charge conserving model. Finally, we find evidence beyond the relaxation time approximation which favors impedances due to reduced quasiparticle coherence (e.g. through high imaginary self-energy or spectral weight shifted away from the Fermi surface) over scattering origins (e.g. domain wall scattering). We stress, however, that further work is necessary to establish the precise microscopic origin of ICM in $NbSe_2$.

**Discussion**

Comprehensive measurements of the temperature, magnetic field and disorder dependence of the MR in $NbSe_2$ confirm the robustness of the LMR and concomitant lack of QOs up to higher field strengths (30 T) and down to lower temperatures (0.35 K) than previously reported. Both phenomena have remained unexplained for 50 years. Our minimal ICM model, adapted to the known FS of $NbSe_2$, is found to offer a coherent explanation for the robust LMR up to the highest field strength, its magnitude, its onset field $H^*$, the vanishing of the MR above $T_{CDW}$, the saturation of the MR in the intermediate temperature range, its disappearance under hydrostatic pressure, the disorder independence of the $H$-linear slope, Kohler scaling, the simultaneous linearity of the Hall effect, and the absence of QOs on all pockets except the selenium pancake pocket. The only failure of the model in its current form is its inability to account for the sign change in the Hall effect, which may result from the lack of FS reconstruction in the model.

As highlighted in the Introduction, multiple scenarios have been proposed to explain LMR and the origin of LMR in a particular metal is likely to depend on the specifics of their electronic state. Before discussing the implications of our modelling for the physics of $NbSe_2$, therefore, we first describe here what aspects of the data are inconsistent with other models for the *H*-linear MR. While the non-linear $\rho_{yx}(H)$ in $NbSe_2$ hints at multiple carriers, no variant of the multi-band Drude model can generate such a robust *H*-linear MR. Models based on random resistor networks (*10*,*49*) or fluctuations in the carrier density (*2*,*19*) can also be ruled out since (i) $\Delta\rho$ is *H*-linear while $\rho_{yx}$ is non-linear, and (ii) at low *T*, the magnitude of the MR is essentially constant while $\rho_{yx}$ changes sign. At the same time, a different disorder dependence would be expected. Previous STM work revealed that CDW domains in $NbSe_2$ shrink with increased disorder down to ~5 nm in strongly disordered crystals (*50*). This is expected to enhance the mobility variance and thereby significantly increase the LMR (*49*,*51*), in contrast with the experimental findings (see Fig. 2). Clearly, additional assumptions or degrees of freedom would be required for the random resistor network picture to capture all the experimental observables captured by the ICM model. The low field scale for the crossover *H*-linear MR (~ 1 T) implies that the quantum model of Abrikosov (*12*) is not relevant here. Moreover, the hotspot model of Koshelev (*20*) predicts $\rho_{yx} \sim H^2$, while the notion that sharp FS corners account for the *H*-linear MR, as proposed for other CDW compounds (*15*) – appears invalid since the robustness of the *H*-linearity up to 30 T implies extremely sharp FS corners and thus reconstruction into smaller pockets, for which one should observe (*13*) QOs and MR saturation well below 30 T in clean $NbSe_2$ – inconsistent with our results. Finally, the robust *H/T* MR scaling behavior seen in several strange (*9*) and quantum critical (*4*,*6*) metals indicates that its origin in those systems is tied to the *T*-linear resistivity. In $NbSe_2$, by contrast, LMR is observed in a regime where the resistivity does not exhibit *T*-linearity.

In the original Peierls picture of CDW order (*52*), nesting of the FS by a wave vector $\vec{Q}$ yields a divergence of the electronic susceptibility. In the presence of even infinitesimal electron-phonon coupling this is accompanied by a sharp dip (Kohn anomaly) in the phonon dispersion causing a combined electronic and lattice instability and reconstruction of the FS (*53*). In this idealized scenario, there is no additional scattering in the hybridized FS after backfolding, beyond the usual isotropic impurity scattering. In practice, perfect nesting does not exist (*54*). In $NbSe_2$ in particular, the applicability of the ideal Peierls picture has long been questioned. It cannot predict the observed CDW vector (*55*), the broad Kohn anomaly (*56*), nor the presence of fluctuations above $T_{CDW}$. Models of the CDW order instead require a momentum dependence of the electron-phonon coupling and inelastic processes away from the Fermi level (*41*,*56*,*57*). Our study suggests that impedances to cyclotron motion are an essential extra ingredient to explain both the LMR and the lack of quantum oscillations. Finally, a scenario based on hotspots formed through strong scattering from (CDW) quantum fluctuations, i.e. due to proximity to a quantum critical point (*4*), seems unlikely here. Thermal variations of the CDW order parameter should be present above $T_{CDW}$, be strongest near $T_{CDW}$ and be suppressed at lower *T* (*33*,*56*). The LMR in $NbSe_2$, meanwhile, is absent above $T_{CDW}$, shows breakdown just below $T_{CDW}$, and is strongest at low *T*.

An alternative type of fluctuation may lie in the strong-coupling nature of the CDW formation (*58*,*59*). Going beyond the idealized Peierls picture, quantitative modeling of the CDW requires renormalizing the bare electrons and phonons with fluctuations beyond the mean field level (*40*,*41*). This results in a strongly momentum-dependent CDW gap, but may equally affect the off-diagonal elements of the gap matrix, which determine the lifetime of renormalized electrons. Under the natural assumption that the momentum dependence of the diagonal and off-diagonal elements track one another, the scattering rate is largest at hotspots along the FS connected by CDW wave vectors.

Another possibility is that impedances are formed by scattering from CDW domain walls. Traversing a sharp domain wall can be thought of as quenching from one CDW structure to one shifted by a single lattice spacing. Bloch's theorem implies that spatially shifting the electronic wave functions adds a momentum-dependent phase factor. For electronic states at hotspots along the FS, the phase shift is guaranteed to yield different electronic states at the FS. The joint density of states available for scattering from CDW domain walls is thus maximized at hotspots, causing their scattering rate to be maximized. Note that although domain sizes larger than the expected cyclotron size at 30 T have been observed in clean $NbSe_2$ (see also SM) (*50*), these do not account for phase slip lines in individual CDW components attached to point defects. The density of such phase slips is determined by the difference between the locally commensurate CDW vector and the incommensurate position of the peak in electronic susceptibility (*40*,*60*,*61*), and may thus be expected to be relatively large and doping-independent.

Estimating the CDW correlation length to be $\xi \leq 50$ nm and using ARPES results for the CDW gap anisotropy (*35*), the inequality $\Delta_{CDW} < \hbar v_F/\xi$ is found to hold for all major FS pockets except for the inner *K* barrel on which the CDW order is strongest. In this regime, it has been suggested that FS reconstruction may be inhibited since the CDW appears glassy to the electrons (*22*). In underdoped cuprates, this inequality is also satisfied and yet QOs of reconstructed Fermi pockets have been observed (*62*,*63*). More work is required to thoroughly test this mechanism. Finally, we point out that two key ARPES studies on $NbSe_2$ (*35*,*44*) show a reduction or even vanishing of the quasiparticle spectral weight at the CDW hotspots. At the same time, the backfolded spectral weight is minimal. Certainly, the origin of the impedances found here appear related to the ARPES results. Neither ARPES nor LMR, however, can by themselves reveal the microscopic origin of the spectral weight reduction nor the impedances. Resolving this will require further investigation.

Overall, our study provides indirect, but robust evidence for the formation of impedances alongside the usual FS reconstruction in the CDW ordered phase of $NbSe_2$. The prototypical nature of $NbSe_2$ among (strongly coupled) density wave systems, both charge and spin, indicates this result may have broader implications. Considering the pnictides (where hotspots from critical spin fluctuations may act as impedances) and optimally doped cuprates (where the edges of Fermi arcs offer a natural source of impedance), the turnover scale of the MR appears entirely insensitive to disorder and scales with $H/T$, indicative of some form of strange metallic transport possibly beyond the relaxation time approximation. Nevertheless, these examples illustrate that impedances naturally emerge across strongly correlated electron systems through the *k*-selectivity of a wide variety of correlation effects with distinctive consequences in the presence of a magnetic field (*21*). Understanding how ICM affects the relatively simple Fermi-liquid $NbSe_2$ is essential if we are to understand related phenomena in more complex unconventional superconductors.

We conclude by arguing that ICM enables a deeper and simpler understanding of the electrical transport and lack of quantum oscillations in $NbSe_2$ than is possible via the conventional FS reconstruction scenario. While further work is required to identify the microscopic origin of ICM in $NbSe_2$, The ICM scenario presented here offers a promising starting point towards a new paradigm for describing density wave order and correlated electron dynamics.

## Materials and Methods

Bulk $2H$-$NbSe_2$ crystals of varying quality, indicated by a varying $T_c$ and *RRR*, were obtained from the crystal provider *HQ Graphene* and cut into bar shapes along the *a*- or *b*-axis using a razor blade. XRD measurements indicated that there were no impurity phases present to within experimental resolution present. We tentatively ascribe the disorder to crystal imperfections (e.g. dislocations), point-like impurities and vacancies created during the growth process. XRD was also used to determine the crystallographic orientation of each crystal. (The crystal structure itself, shown in Fig. 1, was generated using VESTA (*64*)). Electrical contacts were made in a 6-probe Hall bar configuration by fixing 25 μm diameter gold wires to the sides of the samples using 4929N DuPont silver paste, carefully shorting out the *c*-axis to avoid contributions from the out-of-plane resistivity $\rho_c$. Note that the data for the *RRR* = 19 sample in Fig. 2D) is noisier due to a contact that slightly degraded during the experiment. High-field transport properties were measured in a lab-built $^3$He system inserted into a 30 T Bitter magnet at the HFML-FELIX using a Keithley 6221 current source and a Stanford Research SR865 lock-in amplifier. The derivatives shown in Fig. 2 were smoothed with a Savitsky-Golay filter of polynomial order 1. Low-field measurements were performed in a Cryogenics cryogen-free measurement system (CFMS). Modeling was performed using Python code, see SM for details.

**Acknowledgments**

The authors thank C.M. Duffy and J. Ayres for insightful discussions.

**Funding:**

European Research Council (ERC) under the European Union's Horizon 2020 research and innovation program, Grant Agreement No. 835279-Catch-22 (NEH, RDHH)
Engineering and Physical Sciences Research Council EPSRC grant EP/V02986X/1 (NEH)
HFML-FELIX, a member of the European Magnetic Field Laboratory (AK, SW, DP)
Engineering and Physical Sciences Research Council EPSRC (UK) via its membership of the European Magnetic Field Laboratory (grant no. EP/N01085X/1) (NEH)
Engineering and Physical Sciences Research Council EPSRC Grant No. EP/X012239/1 (FF).

**Author contributions:**

Conceptualization: RDHH, NEH
Methodology: AK, DP, SW
Sample orientation: PT
Investigation: AK, DP, SW
Data analysis: AK
Modeling: RDHH
Funding acquisition: NEH, FF
Supervision: NEH
Writing—original draft: AK, RDHH and NEH with key input from FF and JvW

**Competing interests:** Authors declare that they have no competing interests.

**Data, Code, and Materials Availability:** All data and code needed to evaluate and reproduce the results in the paper are present in the paper and/or the Supplementary Materials. The data and code supporting this study can be found at https://doi.org/10.34973/drd0-kj23. All new materials generated by this study are provided in the Materials and Methods.

# Supplementary Materials

## High field dataset

In the main article, we discussed and showed the full temperature ($T$) dependence of the magnetoresistance (MR) of the $NbSe_2$ sample with a residual resistivity ratio $RRR$ = 44. For our disorder study, we measured samples cut from four more crystals with varying $RRR$, whose $R$-$T$ curves are shown in Fig. 1**B** of the main article. All samples were cut along the $a$- or $b$-axis.

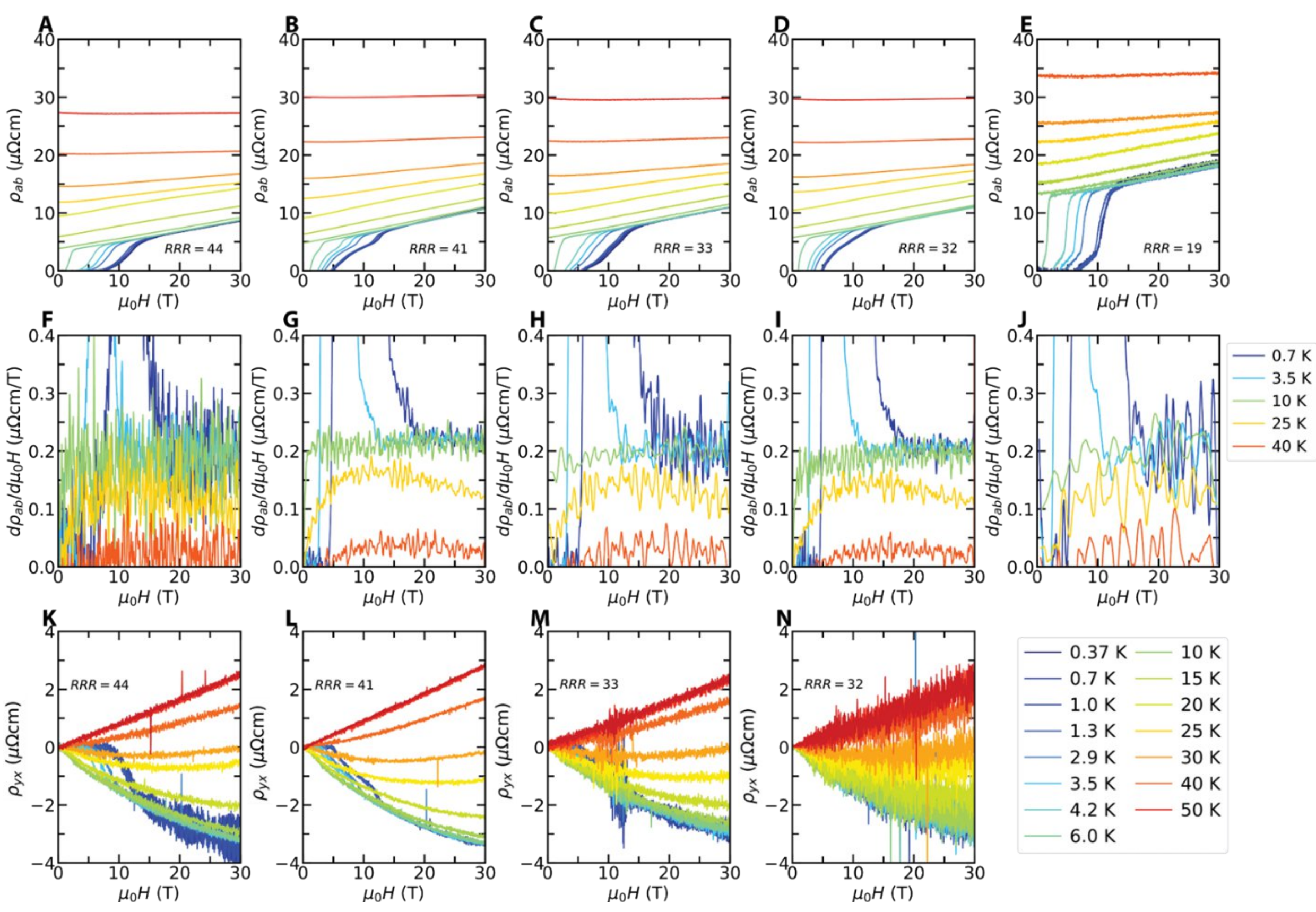

**Fig. S1.** Overview of the full high-field dataset. A)-E) Complete datasets of five samples with $RRR$ values ranging from 19 to 44. All samples show the same qualitative behavior in their MR. Note how the zero field resistivity of all curves broadly increases with decreasing $RRR$ to within experimental error. F)-J) Derivatives of the curves shown in panels A)-E) at selected temperatures. The curves in panels G)-J) have been smoothed using a Savitsky-Golay algorithm with polynomial order 1, while the curves shown in panel F) are raw data. K)-N) Hall resistivity $\rho_{yx}(H)$ measured on the samples with $RRR$ = 44, 41, 33 and 32, respectively. (The Hall contact on the sample with $RRR$ = 19 broke during cooldown.) Note that $R_H$ (=$\rho_{yx}(H)/\mu_0 H$) changes sign at the temperature where the $H$-linearity of $\rho_{yx}$ is lost. The oscillations on the low-$T$ curves in panel I) are due to field-induced vibrations in the setup.

Fig. S1 shows the full $T$-dependence of the MR for all five of these crystals measured between 0.3 K and 50 K. The top five panels show the in-plane resistivity $\rho_{ab}$ as a function of field. All 5 samples show the same qualitative behavior of the MR as the sample discussed in the main article. Furthermore, the absolute value of the resistivity increases with decreasing *RRR*, consistent with Matthiessen's rule. The extra structure at the superconducting phase transition in these samples is attributed to increased inhomogeneity, leading to a variation of $H_{c2}$ inside each sample. Panels **F-J** of Fig S1 show selected first derivatives of the MR curves in order to emphasize the robustness of the $H$-linear slope over a wide temperature and field range, despite a 4-fold change in the residual resistivity $\rho_0$.

Panels **K-M** of Fig S1 show the Hall resistivity $\rho_{yx}$ measured on four of the five samples studied in high field. (There is no data shown for the sample with *RRR* = 19, as it did not have a working Hall contact.) We see that at low $T$, $\rho_{yx}$ exhibits highly non-linear behavior, suggesting that charge carriers from multiple Fermi pockets are participating in the transport. The high-field Hall resistivity is negative and $T$-independent up to 10 K, after which it gradually diminishes before changing sign at $T \approx 25$ K. At higher temperatures, $\rho_{yx}$ remains positive and tends towards linearity, becoming fully $H$-linear above 40 K. We note that the sign change coincides with the breakdown of $H$-linearity in all samples, and the Hall resistivity becoming completely linear coincides with a recovery of a more conventional MR. Curiously, the sign change in the Hall coefficient $R_H$ happens for all samples at roughly 2/3 $T_{CDW}$, where $T_{CDW}$ is the transition temperature into the charge density wave (CDW) state that changes slightly under the influence of disorder. As $R_H$ changes sign at the same temperature where $H$-linearity is lost, we interpret the sign change as a consequence of the weakening of CDW order. Because of this, the impedances on the Fermi surface (FS) weaken, and we slowly recover the high-$T$ hole-like FS of $NbSe_2$. Our data suggest that the impedances are caused by the CDW order which strengthens substantially down to 15 K.

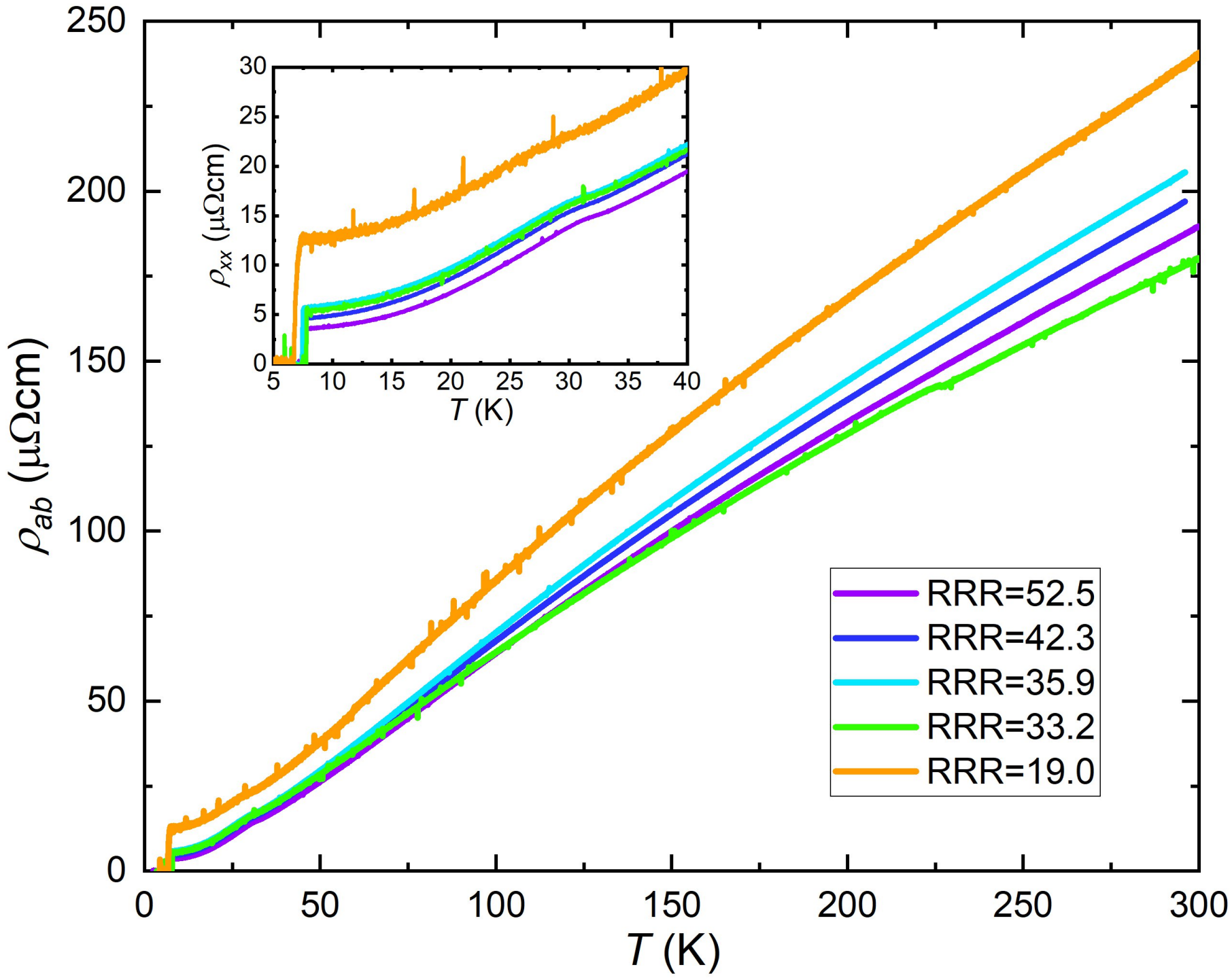

**Fig. S2**. $\rho_{ab}(T)$ curves of the $NbSe_2$ samples studied at low field. Cooldown curves of the five samples studied in low field in the CFMS. The samples with *RRR* = 52, 42 and 36 were cut from the same mother crystals as the high-field samples with *RRR* = 44, 41 and 33, respectively. The samples with *RRR* = 33 and *RRR* =19 are the same samples as the high-field samples with *RRR* = 33.2 and *RRR* = 19.4, but with fresh electrical contacts. We see that at high temperatures, Matthiessen's rule is violated, while below $T_{CDW}$, it is recovered. Inset: Zoom in at low $T$ showing Matthiessen's rule is obeyed at low temperature.

**Low-field dataset**

To assess the dependence of the crossover field $\mu_0 H^*$ on the zero-field resistivity $\rho_{ab}(0)$, and to more closely map the breakdown of *H*-linearity upon increasing temperature, we performed a set of low field measurements up to 8 T in an automated cryogen-free measurement system (CFMS) from *Cryogenics Ltd*. For this study, we reused the samples with *RRR* = 33 and *RRR* = 19, but with new electrical contacts. For the other three batches, from which the samples with *RRR* = 44, 41 and 32 originate, three new samples were cut from the same mother crystals. The new samples have *RRR* values of 52, 42 and 36, respectively. The low field data were taken at intervals of 1 K or 2 K. All samples were contacted in a Hall bar geometry to allow for simultaneous measurements of $\rho_{ab}$ and $\rho_{yx}$. All samples were again cut along the *a*- or *b*-axis.

Fig. S2 shows $\rho_{ab}(T)$ of these 5 samples. We again see that Matthiessen's rule is satisfied at low *T*, and that we have a variation in the residual resistivity $\rho_0$ by a factor of around 4. The field-dependent data obtained with the CFMS are shown in Fig. S3. In panels **A**-**E** of Fig. S3, we show $\rho_{ab}(H)$ for all five measured samples. All curves were measured between 6 K and 36 K. To within experimental error, the measured resistivities are consistent with the high-field data reported in Fig. S1. In panels **F**-**J** of Fig. S3, we show the corresponding field derivatives. From these curves, we can see that the crossover scale $\mu_0 H^*$, defined as the field where $d\rho_{ab}/d\mu_0 H$ reaches 90% of its high field value, is always around 1-2 T, and increases with increasing $\rho_{ab}(0)$. Furthermore, we can clearly see the gradual breakdown of the *H*-linearity with increasing temperature, which typically occurs around *T* = 18 K, after which the slope also diminishes. Hence, above this temperature scale, CDW correlations are sufficiently suppressed to allow electrons to tunnel through the impedances.

Panels **K**-**O** of Fig. S3 show the Hall resistivity measured on the 5 samples with varying *RRR* values up to 8 T. Note that these curves have been smoothed using a Gaussian smoothing algorithm. The Hall resistivity $\rho_{yx}$ is temperature independent up to 15 K, after which it slowly starts to decrease in magnitude. Around 25 K, $R_H$ changes sign, indicated by a $\rho_{yx}(H)$ which is zero over almost the entire field range. For $T > 25$ K, $\rho_{yx}$ is positive and becomes progressively more linear. Above $T_{CDW}$, $\rho_{yx}$ becomes fully *H*-linear and independent of temperature within the studied temperature range. Again, the magnitude of the Hall effect is consistent with the high field data to within experimental error.

We see here that the sign change in $R_H$ and the breakdown of *H*-linearity do not occur at the exact same temperature. Nevertheless, we can still ascribe the sign change in $R_H$ to a weakening of the CDW order. In panels **F**-**J** of Fig. S3, we see that although *H*-linearity is lost above 18 K, there is still a sizable MR between 18 K and 25 K. We know from the results of the modelling that the conventional orbital MR of the high-*T* FS of $NbSe_2$ is small compared to the magnitude of the observed LMR, suggesting the presence of a remnant contribution from ICM in this temperature range. Furthermore, 18 K coincides with the temperature where $R_H$ starts to strongly deviate from its low-*T* value. Therefore, the temperature at which $R_H$ changes sign appears to coincide with the temperature at which tunneling through the impedances on the FS caused by the CDW order is possible. However, the Hall effect appears more sensitive to such breakdown than the MR. ICM shows a similar dependence on FS topology as the Hall effect

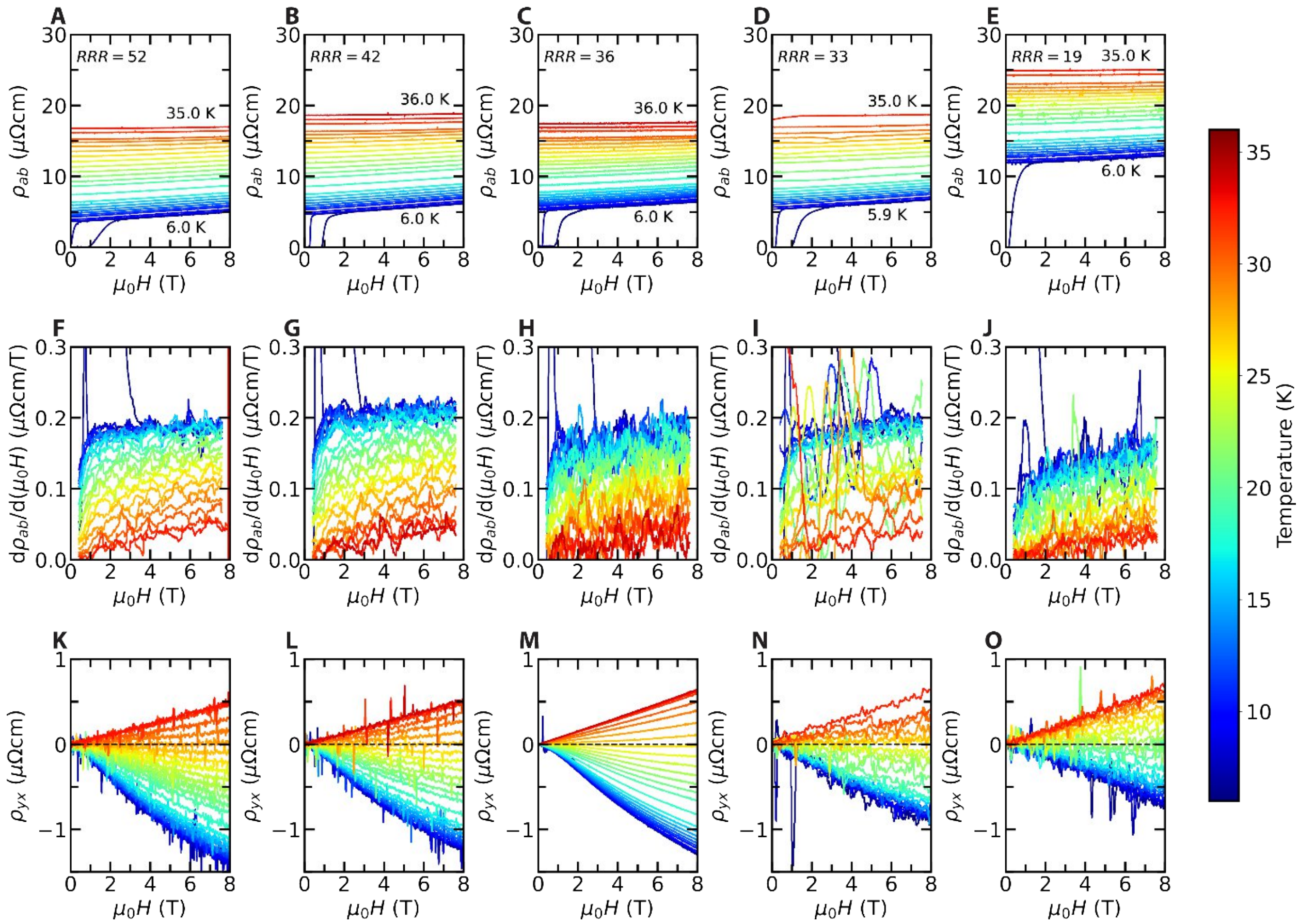

**Fig. S3**. Overview of the low-field data set. The samples with *RRR* = 52, 42 and 36 were cut from the same mother crystals as the high-field samples with *RRR* = 44, 41 and 33, respectively. The samples with *RRR* = 33 and *RRR* = 19 are the same samples as the high-field samples with *RRR* = 33.2 and *RRR* = 19.4, but with fresh electrical contacts. All curves have been measured in temperature steps of 1 K or 2 K. A)-E) $\rho_{ab}$ as a function of field up to 8 T for the five measured samples. Note a gradual increase of the zero-field resistivity for increasing *RRR*. The minimum and maximum temperature at which a field sweep was performed have been indicated in the figure. F)-J) Derivatives of the curves shown in panels A)-E). These curves clearly show the breakdown of *H*-linearity for temperatures above 18 K. The fluctuations in I are due to equipment issues. K)-O) The Hall resistivity measured on the five crystals. The overall shape of the Hall effect is not too affected by disorder, but the magnitude at lower temperatures does seem to be affected. Note that the Hall effect changes sign, indicated by a near-zero $\rho_{yx}$, exactly at the temperature where *H*-linearity breaks down. The $\rho_{yx}(H)$ curves have been smoothed using a Gaussian filter.

**Longitudinal magnetoresistance**

The concept of impeded cyclotron motion assumes that the MR is predominantly orbital in nature. If this is the case, the MR is expected to vanish for a field parallel to the current, since in this configuration the charge carriers do not experience a Lorentz force. To test this, we also measured the MR of the high-field samples with **H**//I//*ab* at *T* = 0.37 K and 4.2 K (deep within the ICM regime) and at *T* = 30 K, i.e. close to where the CDW vanishes and ICM fully breaks down.

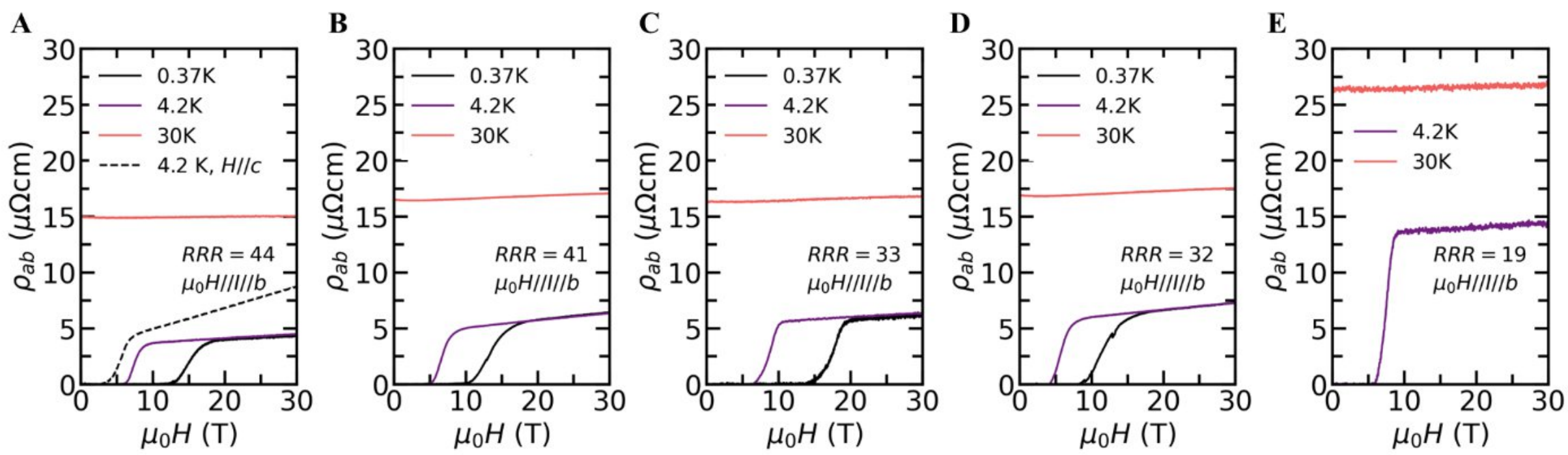

**Fig. S4**. Evidence for the orbital nature of the low-*T* MR in $NbSe_2$. A)-E) In-plane resistivity as a function of magnetic field for selected temperatures, with the field oriented parallel to the current. In this configuration, we observe a significant reduction in the magnitude of the MR, implying that the MR presented in the main article is predominantly of orbital character. This disfavors a non-orbital contribution to the MR as a mechanism for the observed MR profile. In panel A), the 4.2 K curve for $\vec{H}//c$ has been included for comparison.

Figure S4 shows the resultant data. A finite MR is still visible in all samples, although its magnitude is much reduced compared to the transverse MR (**H**//*c*). This nonzero MR most likely arises from a misorientation of the samples in the magnetic field. The angle with respect to the applied field can be minimized by monitoring the signal of a Hall bar as the probe is rotated, though a small, residual angle (~3°) between the sample orientation and the Hall bar is sufficient to explain the observed MR. We therefore conclude that the low-*T* MR response in $NbSe_2$ is indeed orbital in nature.

**Quantum oscillations**

A striking feature in the MR of $NbSe_2$ is the absence of quantum oscillations (QOs) from any of its major pockets even in the cleanest samples (*31*,*32*). These are expected at frequencies of 5-10 kT without and 0.05-2 kT with FS reconstruction (see Table S1 below). Indeed, no QOs are visible in any of the data shown in Fig. 2, Fig. S1 and Fig S3. The only pocket known to exhibit QOs (through the de Haas-van Alphen or dHvA effect), is the pancake-like pocket centered around the Γ-point in the Brillouin zone (BZ). Here we show that we also observe, for the first time, Shubnikov-de Haas oscillations arising from this pocket, i.e. in the electrical resistivity, in a sample with *RRR* = 39. Note that on this sample, the *c*-axis was not properly shorted out, so the *c*-axis possibly also contributes to the transport. For this reason, we report here only resistance data, not resistivity data.

Fig. S5**A** shows MR curves *R*(*B*) between 0.9 K and 18 K. We again see a robust *H*-linear MR that spans a wide temperature and field range. Furthermore, we see no QOs down to the lowest temperatures. In Fig. S5**B** we show the angle dependence of *R*(*B*) versus field measured at 1.35 K, where the angle $\theta$ is defined as 0° for a magnetic field oriented along the *c*-axis, and $\theta$=90° for the field parallel to the current. When we rotate the field away from the *c*-axis, we observe an increase in the MR up to roughly 40°, after which the MR decreases in magnitude. Furthermore, for angles larger than 50°, we observe the onset of QOs which increase in magnitude with increasing angle. We plot the derivative d*R*/d(1/*B*) versus 1/*B* in Fig. S5**C** to remove the background from the QOs. In the inset, we plot the angle dependence of the QO frequency, which shows a clear decrease with increasing angle $\theta$. These angular dependences are representative of a strongly ellipsoidal orbit in the *ab*-plane whose semi-minor axis grows rapidly when the field is rotated out of the plane. This confirms that we are observing QOs from the Se pancake-like pocket at the Γ-point of the first BZ, consistent with the observations in Ref. (*32*). This implies that the absence of QOs from the pockets participating in the CDW order is not a matter of sample quality, but rather has a more intrinsic physical mechanism behind it.

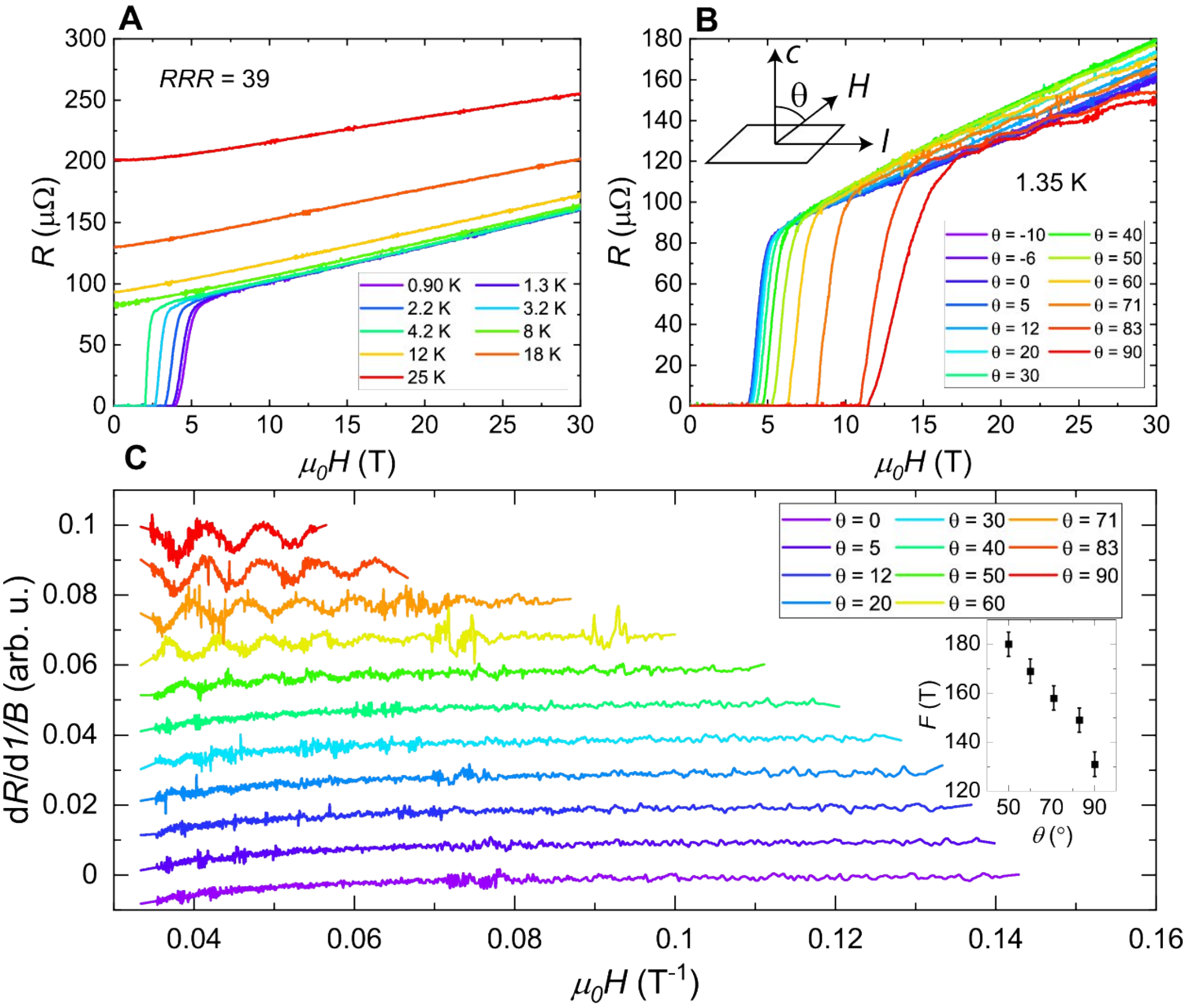

**Fig. S5**. Angle dependence of a sample with *RRR* = 39. A) Resistance *R* as a function of out-of-plane magnetic field. This sample was only measured in the temperature range where the MR is linear. B) Angle dependence of the resistance of $NbSe_2$ at *T* = 1.35 K. The angle *θ* is defined with respect to the *c*-axis, as shown in the inset. The MR remains linear up to *θ*=40°, although the slope gradually increases. For *θ* > 50°, the MR quickly diminishes and becomes non-linear. At this angle, we also see an onset of QOs. C) Waterfall plot of the QOs. We observe no QOs for *θ* < 50°. Beyond *θ* = 50°, we observe QOs that increase in amplitude and decrease in frequency, suggesting that they originate from the pancake-like pocket at the Γ-point in the BZ as reported in Ref. (*32*). Curves have been offset for clarity. Inset shows the QO frequency as a function of *θ*. Note that the electrical contacts for this sample were mounted differently to the other crystals in this study. Here, contacts were attached onto the top surface in a Van der Pauw geometry, and the *c*-axis was not shorted. As a result, there is a contribution from $\rho_c$ in this data set that likely contributes to the non-monotonic behaviour of the MR with respect to angle.

**Conductivity models**

Below we report on a variety of theoretical models with direct relevance to the origin of LMR in $NbSe_2$. In section A, we discuss the conductivity formula and form for the hotspots used for the ICM model in the main text. We detail the insensitivity of the model to the width and strength of hotspots at low $T$ and our approach to incorporate magnetic breakdown just below $T_{CDW}$.

In section B, we explore the expectations within the Peierls picture. First, we establish the reconstructed FS itself as well as the myriad of expected QOs. Secondly, we evaluate the conductivity to arbitrary magnetic field strength on this complex FS in both the isotropic lifetime and isotropic-$\ell$ regimes. We find neither version can explain the extended $H$-linear regime with robust slope nor absence of quantum oscillations in experiment.

Finally, in section C we explore effects beyond the relaxation time approximation (RTA). This requires introducing a new framework expanding on previous work by the present authors in quasi-1D (*48*). We use this framework to show 1) small-angle scattering reduces the $H$-linear slope, 2) present an impeded cyclotron motion model with strict charge conservation and 3) take into account the non-colinear velocity at hotspots. The latter results suggest that termination of quasiparticle coherence at hotspots (e.g. through large imaginary self-energy or removal of spectral weight away from the Fermi level) explains the data better than scattering hotspots (e.g. by domain walls). We emphasize that none of the extra models presented here substantially improve upon the simple, phenomenological model presented in the main text. They serve to confirm the necessity of impeded cyclotron motion and the robustness of the results to a wide range of corrections.

A. Relaxation time model of impeded cyclotron motion

We now detail the model used to create the theory results shown in the main article. In order to model the impedances to cyclotron motion, we use the most general solution to the Boltzmann transport equation within the RTA and neglect thermal broadening. The resulting general formula was also used to propose impedances to cyclotron motion in Ref. (*21*) and is known as the Shockley-Chambers Tube Integral Formalism (SCTIF) (*45*,*46*):

$$\sigma_{ij} = \frac{e^2}{\pi\hbar}\int_{\mathrm{FS}} \frac{\mathrm{d}^2\vec{k}}{(2\pi)^2}\frac{v_i}{v}\int_0^\infty \mathrm{d}t v_j(-t)\exp\left(-\int_0^t \frac{\mathrm{d}t'}{\tau\left(-t'\right)}\right) \tag{S1}$$

Here, $\sigma$ is the conductivity, indices $i$,$j$ are $x$,$y$ (since the $k_z$ corrugation of the FS is neglected), $e$ is the electron charge and positive, $\hbar$ the reduced Planck's constant, FS refers to the FS including a sum over the different pockets, $\vec{v}$ is the Fermi velocity with implicit $\vec{k}$-dependence and $\tau$ the scattering rate in the RTA again with implicit $\vec{k}$-dependence. Spin degeneracy is included in the formula. The exponent in Eq. (S1) is the probability for a quasiparticle to survive to time $t$ under changing scattering lifetime as the quasiparticle traverses the FS. Time evolves the position of a quasiparticle on the FS due to cyclotron orbital motion $\vec{k}(t)$ and thereby changes both $\vec{v}$ and $\tau$. The integral is evaluated without approximation for the strength of the magnetic field. The velocity is defined through the energy dispersion as usual via

$$\vec{v} := \frac{1}{\hbar}\vec{\nabla}_k\varepsilon_k \tag{S2}$$

The energy dispersion used in the calculations is the tight-binding expansion of Ref. (*35*) based on detailed ARPES results. This tight-binding parameterization satisfies charge conservation, meaning that the combined bands contain one hole per unit cell. The pancake pocket at the Γ-point is neglected throughout for it only contributes significantly to the $z$-component of the conductivity.

The final ingredient to fully define Eq. (S1) is the scattering lifetime across the FS. We pragmatically choose the form $\tau(\vec{k}) = \tau_c + \alpha(\vec{k})(\tau_h - \tau_c)$. Here, $\tau_c$ is the isotropic cold scattering time and the main degree of freedom of the model. $\tau_c$ is fit to the zero-field resistivity in experiment and varies between 0.1 and 0.3 ps, dependent on temperature. LMR is reached when cold charge can traverse between subsequent hot spots. Thus, $\tau_c$ is inversely proportional to the turnover scale of the LMR. For $\tau_h$ we seek a form which: 1) is local in $k$-space; 2) has no discontinuity; 3) is nevertheless sufficiently strong to impede cyclotron motion: 4) contains minimal degrees of freedom. We note that once the probability for charge to traverse a hotspot is below ~1% (in order to prevent magnetic breakdown), the absolute magnitude of $\tau_h$ is irrelevant. The difference between the addition of scattering times or rates is similarly irrelevant to the outcome of the model. We require a smooth $\tau_h$ rather than a delta function such that a strictly $H^2$ MR remains at zero field as also found in Ref. (*21*). In the end, charge at the hotspots themselves are shorted out (contribute negligibly to the conductivity) and the MR is generated by the cold charge subject to essentially isotropic scattering $\tau_c$ driven into the impedances. We use a Gaussian form for $\alpha(\vec{k})$ to satisfy these requirements with minimal degrees of freedom.

$$\tau\left(\vec{k}\right) = \tau_c + \alpha\left(\vec{k}\right)\left(\tau_h - \tau_c\right) \tag{S3}$$

$$\alpha\left(\vec{k}\right) = \exp\left(-\frac{min\left(\varepsilon^2_{\vec{k}+\vec{Q}}\right)}{2\varepsilon^2_{\Delta}}\right) \tag{S4}$$

Here, $\varepsilon_{\vec{k}+\vec{Q}}$ is the energy after translation by $\vec{Q}$ where $\varepsilon_{\vec{k}} = 0$ at the Fermi level. The minimum then selects the value closest to the Fermi level from the 12 different options (3 unique $\vec{Q}$ as well as $\pm\vec{Q}$ for each and 2 bands). The central result in the main article further selects among these options in order to only allow for intra-pocket connections within individual bands. This choice is dictated by ARPES results in Refs. (*35*,*44*), which both find spectral weight reduction only where intra-pocket connections are formed by the CDW order. Later theoretical work (*41*) explained also that interband phonon coupling is expected to be negligible. We have neglected the mass enhancement factor measured in (*35*) for the angle dependence does not cover the full Fermi surface and lacks a temperature dependence to perform consistent calculations. However, we do not expect this to have meaningful impact on the results. The reason is that our model cannot differentiate lifetime or mass individually, only their ratio $\tau/m^*$ -- i.e. the mean free path. Thus, the isotropic mass enhancement is naturally absorbed into the cold scattering rate when we fit $\rho_{ab}(H=0)$. Meanwhile, the effective mass peaks are absorbed into the hot spots. Details about this procedure can be found in Ref. (*65*). The shape of the hotspot and dependence of the result on the value of $\tau_h$ is shown in Fig. S6. The resulting shape of the hotspot used to model the data at $T$ = 10 K and 25 K for the $RRR$ = 44 sample is shown in Fig. 3**D** of the main article. As shown in Fig. S6, above a certain hotspot strength the probability for charge to survive the hotspot is irrelevant within the experimentally accessible field range. Above this threshold (below ~1% probability) the strength of the hotspot is irrelevant and cyclotron motion is fully impeded. In this regime, $\tau_h$ may well vanish and is therefore no longer a fitting parameter. However, when the hotspots are weakened such that breakdown is possible, the MR saturates and the saturation field scale sensitively depends on $\tau_h$, which is the case for the 25 K model data shown in Fig. 3**B** of the main text.

The parameter $\epsilon_\Delta$ controls the width of the hotspots. At high values (broad hotspots, $\epsilon_\Delta > 10$ meV) we find a noticeable increase in the turnover scale of the MR, see Fig. S7. Note that this trend is the opposite of the single-band single-hotspot model presented in Ref. (*21*), where increasing the hotspot width results in a small decrease in $H^*$ due to the reduced distance between subsequent impedances. This difference is a consequence of the multiband character of $NbSe_2$ and the presence of variable inter-hotspot distances.

It is important to stress that the model uses the fewest possible degrees of freedom to model the formation of impedances on the FS of $NbSe_2$ while using the most general solution to the Boltzmann transport equation within the RTA. The ingredients for the model are the FS shape (determined by the tight-binding coefficients from ARPES (*35*)), locations of the impedances (determined by the known $\vec{Q}$), the cold scattering lifetime $\tau_c$ (fit to the zero-field resistivity), the width of the impedances (irrelevant as long as magnetic breakdown is absent) and the strength of the impedances (irrelevant as long as magnetic breakdown is absent). Note that among these ingredients, only one ($\tau_c$) constitutes a variable fitting parameter at $T$ < 15 K.

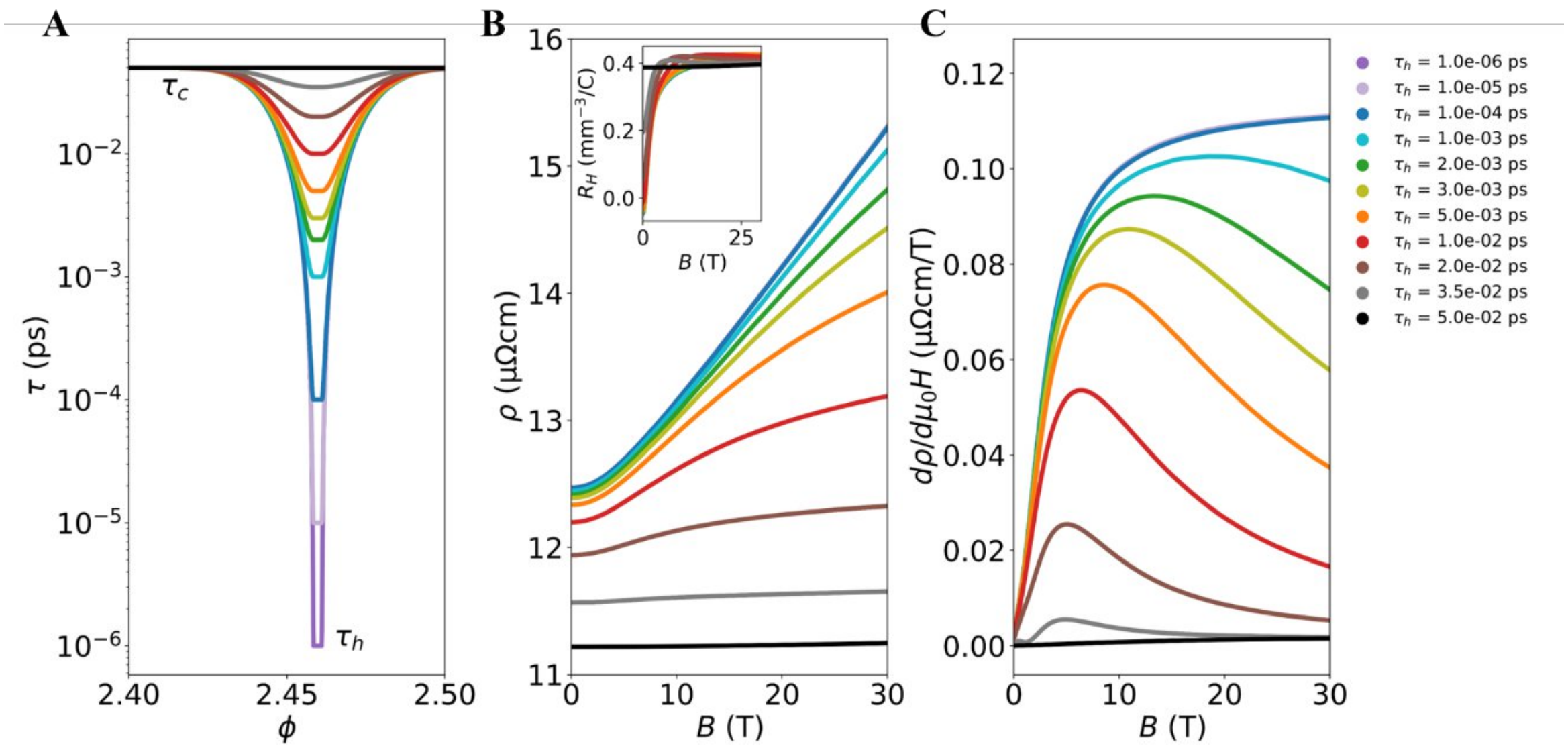

**Fig. S6**. Hot scattering rate dependence of the model and breakdown. A) Representative shape of a hotspot on the inner Γ pocket defined by Eq. (S3)-(S4) at fixed Δ = 10 meV and $\boldsymbol{\tau_c} =$ 0.05 ps for *T* = 25 K. The change in zero-field resistivity results from the fraction of the Fermi surface covered by hotspots. B) Resistivity as a function of hotspot strength. The inset shows that the hotspots leave the high-field Hall effect largely unchanged from the isotropic-τ result and is hole-like (black line). C) Corresponding derivative of the MR. Above a critical value, the hotspot strength is irrelevant in the fully formed ICM regime. As the hotspot is weakened, charge develops a finite probability for breakdown. This manifests in the MR through saturation which matches the shape and field range observed experimentally (see Fig. 3 of the main text or Fig. S1).

Eq. (S1) is evaluated without further approximations. In particular, $\vec{v}$ and $\tau$ update continuously along the full cyclotron orbit $\vec{k}(t)$. This holds true in the zero-field limit when no cyclotron motion occurs all the way to the high-field regime beyond magnetic breakdown when cyclotron motion spans multiple orbits and $\tau$ changes by orders of magnitude repeatedly. Computationally, this poses a challenge. A custom-made adaptive Simpson integration routine was developed for the *t* integral in time-ordered fashion. Using time-ordered integration, the probability for a quasiparticle to survive (the exponential term in Eq. (S1)) can be updated at a computationally low cost and the timestep dynamically updated. Note that although we chose to compute the results in the time basis, formally transforming these integrals to $\varphi$ using d$\varphi$ = $\omega_c$d$t$ does not change the result. The code is in quantitative agreement with Drude theory under a magnetic field and reproduces the results in Ref. (*21*) to numerical accuracy (better than 1 in $10^6$). The resistivity is obtained a full matrix inversion of the conductivity tensor for all calculations. Further details and code are available in Ref. (*65*).

In the main article, we only show fitting at *T* = 10 K (with fully-formed impedances and 1 fitting parameter, $\tau_c$) and for *T* = 25 K (with breakdown and 3 parameters: $\tau_c$, $\tau_h$ and $\epsilon_\Delta$). While these minimal parameterizations led to a fit with a 30% deviation from the measured LMR slope, they nevertheless highlight the fact that with increasing *T*, the impedances will likely weaken and lead to magnetic breakdown.

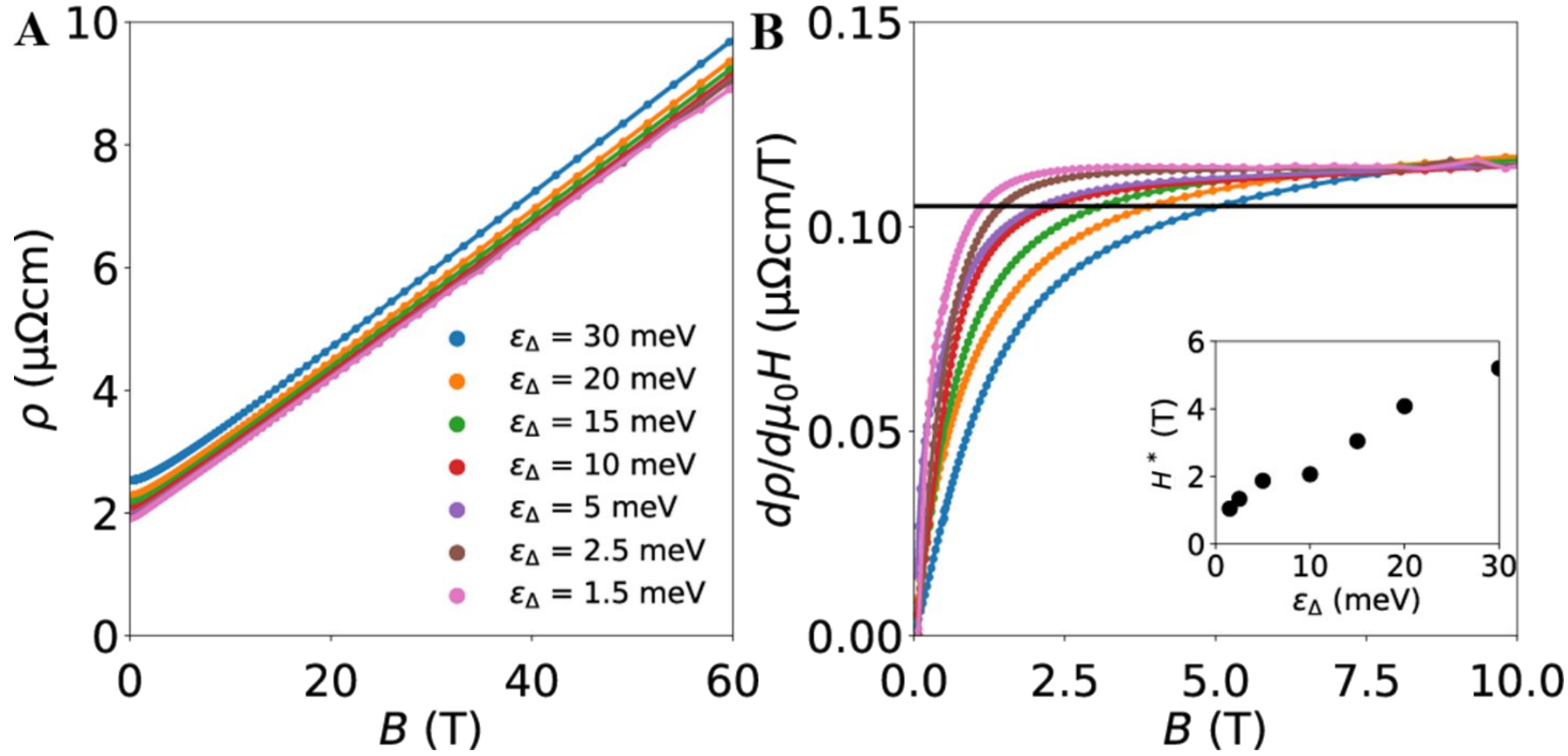

**Fig. S7**. Dependence of the model on $\varepsilon_\Delta$. A) Resistivity as a function of Δ, where higher values correspond to wider hotspots. The change in zero-field resistivity is the consequence of the change in cold carrier density. B) Derivative of the data in panel A) in the low-field regime. The data show the *H*-linear MR is barely affected while the turnover scale increases with increasing $\varepsilon_\Delta$. The inset shows quantitatively the evolution of $H^*$, where the MR reaches 90% of the high-field *H*-linear slope (indicated by the horizonal black line in the main panel). $\tau_c$ = 3 ps is used to obtain a realistic zero-field resistivity for experiment at low *T*, while $\tau_h$ is kept sufficiently low that no magnetic breakdown is possible.

To gain more detailed insight, we consider here a more precise fit to the data. The full temperature range for the *RRR* = 41 sample (with the highest signal to noise ratio) is shown in Fig. S8. In order to produce a precise fit, we resort to renormalizing our observed resistivity values to fit the *H*-linear slope of the model, since the model itself has no degrees of freedom. We find that a good fit is obtained by renormalizing to a room temperature resistivity value to 110 $\mu\Omega\text{cm}$. This is a significant change of about 40 %, but the spread of experimental resistivity data is large due to geometric uncertainty and the quoted room temperature resistivity is in accordance with Ref. (*37*).

After this renormalization, we fit the data as shown in Fig. S8 with parameters listed in Table S2. As the onset of breakdown suggests, we find that the hotspots weaken with increasing *T*. Indeed, $\tau_c$ and $\tau_h$ converge close to $T_{\text{CDW}}$. Meanwhile, we find that the hotspots broaden substantially. This quantitatively explains the low $H^2$ coefficient (high turnover scale $H^*$) at low magnetic fields at *T* = 25 and 30 K (deviating from *T*-dependent Kohler's rule). Thus, $\epsilon_\Delta$ grows large towards $T_{\text{CDW}}$, an opposite *T*-dependence to what is expected of an order parameter. Instead, we find this trend resonates with the broad softening of phonon modes near $Q_{\text{CDW}}$ (as opposed to a sharp Kohn anomaly) observed experimentally in Ref. (*56*). Our results indicate that below $T_{\text{CDW}}$, $Q_{\text{CDW}}$ gradually sharpens or locks in, reducing the size of hotspots (i.e. $\epsilon_\Delta$). Such a description suggests that the coherence length of the phonons relevant in the strong coupling limit is the key driver to hotspot size rather than the CDW gap itself.

The nominal disorder dependence of the model follows Kohler's rule by changing $\tau_c$ to match changes in the residual resistivity. Indeed, this is what is observed experimentally as shown in Fig. 3C of the main article and Fig. S8E-F). We note that the next-leading-order effect of disorder is expected to be a reduction in phonon lifetimes, manifesting as a wider $Q_{\text{CDW}}$ creating broader hotspots. In our model, $\epsilon_\Delta$ would increase in disordered crystals leading to a higher turnover scale $H^*$ in experiment (see Fig. S7). Given the reduction in $T_{\text{CDW}}$ between our clean and most disordered crystal of 10 %, we must conclude that the experimental data in Fig. S8E) is not accurate enough to make any conclusions on weak deviations from Kohler scaling of this sort.

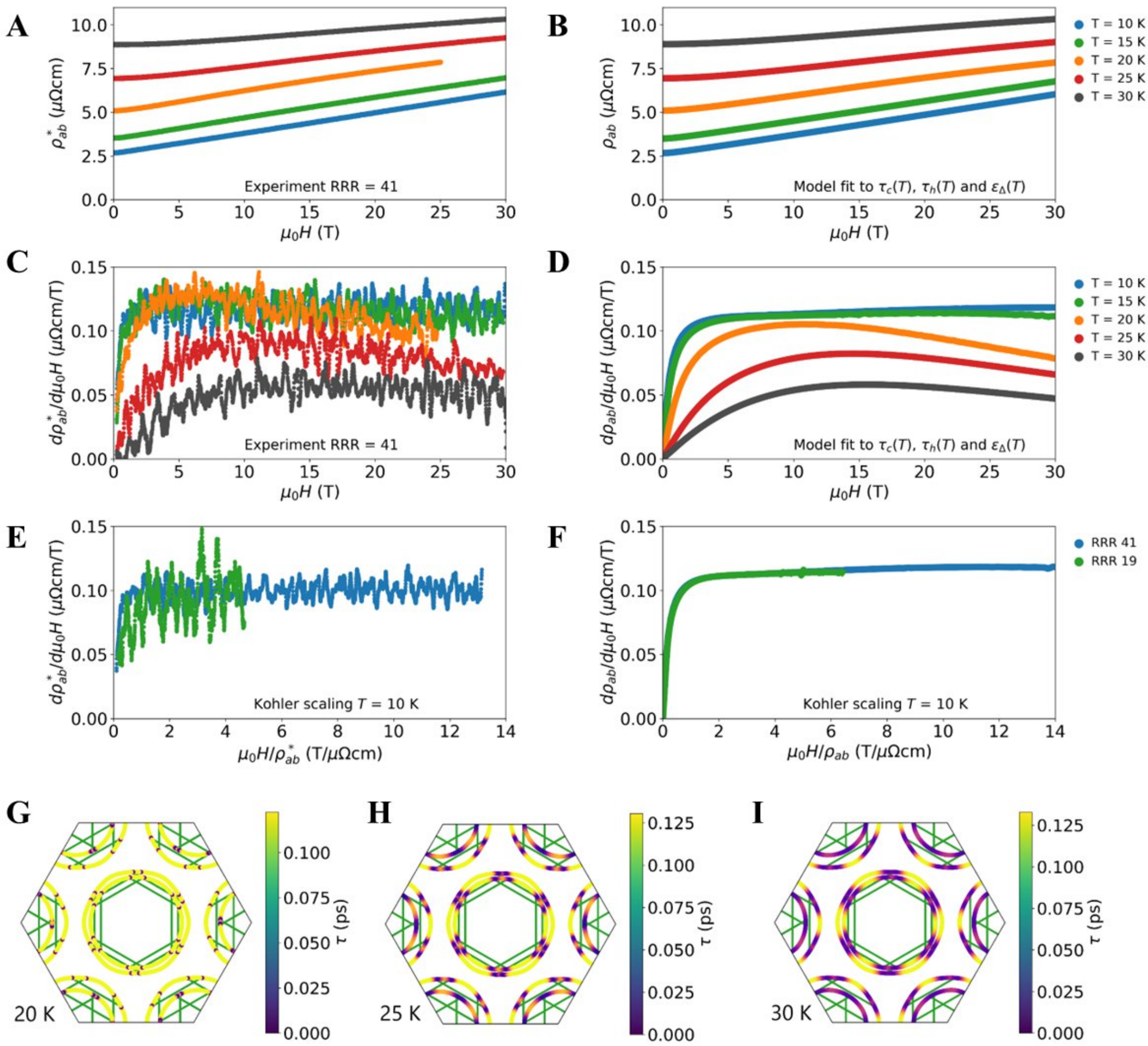

**Fig. S8:** Fit to the full (T, B, disorder) dependence of the MR of $NbSe_2$. A) Resistivity data for the *RRR* = 41 sample with the highest signal to noise. Note here that the experimental data have been renormalized by 40 % in order to match the room temperature resistivity value of 110 μΩcm reported in Ref. (37) and the LMR slope of the model. Note too that the 20 K sweep experienced temperature drift above 25 T and as a result, those data are not shown. B) Corresponding fit using all degrees of freedom in the model, i.e. $\tau_c$, $\tau_h$ and $\epsilon_\Delta$. Values are listed in Table S2. C) Derivative of panel A). D) Derivative of panel B) showing a precise fit of not only the size but also the H-dependence of the MR in absolute units. E) Comparison with the most disordered sample at *T* = 10 K. F) The model nominally only changes $\tau_c$ due to disorder and satisfies Kohler scaling as shown. If the CDW hotspots broaden with increased disorder, this would result in a higher turnover scale in the model, but such an effect could not be resolved experimentally. G-I) Scattering rate of the fitted model at temperatures where breakdown is observed and where the hotspot is neither arbitrarily strong nor wide. We find that the low $H^2$ MR at elevated T can only be captured in the model by substantially broadening the hotspots on approach to $T_{CDW}$.

B. Relaxation time models incorporating the reconstruction

The Peierls reconstruction model forms the basis of our understanding of CDW order. In this section, we present in-depth calculations which include this reconstruction. To the best of our knowledge, such an endeavor has not been attempted before. The key finding is that the Peierls reconstruction alone is insufficient to explain the experimental results. In particular, these calculations show that without lifetime effects in addition to the standard Peierls model 1) a robust non-saturating *H*-linear MR cannot be obtained, and 2) QOs are expected to appear under the experimental conditions. Only the sign change of the Hall effect is captured.

We start by implementing the CDW reconstruction on the ARPES-derived FS of $NbSe_2$ (*35*). To proceed, the pancake pocket as well as the small incommensuration of the CDW order are neglected. Practically, we fold back the FS with $\vec{Q}_1$ = ⅔ΓM as well as $\vec{Q}_2$ and $\vec{Q}_3$ obtained by a 120°and 240° counter-clockwise rotation of $\vec{Q}_1$, respectively. Given that $3\vec{Q}$ = Γ, we also fold using $2\vec{Q}_1 = -\vec{Q}_1$, $-\vec{Q}_2$ and $-\vec{Q}_3$. In order to obtain a closed group of *Q*-vectors under addition, we must also include $\vec{Q}_4 = \vec{Q}_1 + \vec{Q}_2$ and $-\vec{Q}_4$. In total this adds up to nine copies of the FS for each of the two bands, resulting in an 18×18 size matrix to construct the Hamiltonian. Copies which are separated by $\pm\vec{Q}_1$, $\pm\vec{Q}_2$ or $\pm\vec{Q}_3$ open an intraband (Δ) or interband (Ω) gap, while higher-order couplings are neglected. The final Hamiltonian is as follows:

$$\hat{H} = \begin{pmatrix}
\varepsilon^1_{\vec{0}} & \Delta & \Delta & \Delta & \Delta & \Delta & \Delta & 0 & 0 & 0 & \Omega & \Omega & \Omega & \Omega & \Omega & \Omega & 0 & 0 \\
\Delta & \varepsilon^1_{\vec{Q}_1} & \Delta & \Delta & 0 & 0 & \Delta & \Delta & \Delta & \Omega & 0 & \Omega & \Omega & 0 & 0 & \Omega & \Omega & \Omega \\
\Delta & \Delta & \varepsilon^1_{-\vec{Q}_1} & 0 & \Delta & \Delta & 0 & \Delta & \Delta & \Omega & \Omega & 0 & 0 & \Omega & \Omega & 0 & \Omega & \Omega \\
\Delta & \Delta & 0 & \varepsilon^1_{\vec{Q}_2} & \Delta & \Delta & 0 & \Delta & \Delta & \Omega & \Omega & 0 & 0 & \Omega & \Omega & 0 & \Omega & \Omega \\
\Delta & 0 & \Delta & \Delta & \varepsilon^1_{-\vec{Q}_2} & 0 & \Delta & \Delta & \Delta & \Omega & 0 & \Omega & \Omega & 0 & 0 & \Omega & \Omega & \Omega \\
\Delta & 0 & \Delta & \Delta & 0 & \varepsilon^1_{\vec{Q}_3} & \Delta & \Delta & \Delta & \Omega & 0 & \Omega & \Omega & 0 & 0 & \Omega & \Omega & \Omega \\
\Delta & \Delta & 0 & 0 & \Delta & \Delta & \varepsilon^1_{-\vec{Q}_3} & \Delta & \Delta & \Omega & \Omega & 0 & 0 & \Omega & \Omega & 0 & \Omega & \Omega \\
0 & \Delta & \Delta & \Delta & \Delta & \Delta & \Delta & \varepsilon^1_{\vec{Q}_4} & 0 & 0 & \Omega & \Omega & \Omega & \Omega & \Omega & \Omega & 0 & 0 \\
0 & \Delta & \Delta & \Delta & \Delta & \Delta & \Delta & 0 & \varepsilon^1_{-\vec{Q}_4} & 0 & \Omega & \Omega & \Omega & \Omega & \Omega & \Omega & 0 & 0 \\
0 & \Omega & \Omega & \Omega & \Omega & \Omega & \Omega & 0 & 0 & \varepsilon^2_{\vec{0}} & \Delta & \Delta & \Delta & \Delta & \Delta & \Delta & 0 & 0 \\
\Omega & 0 & \Omega & \Omega & 0 & 0 & \Omega & \Omega & \Omega & \Delta & \varepsilon^2_{\vec{Q}_1} & \Delta & \Delta & 0 & 0 & \Delta & \Delta & \Delta \\
\Omega & \Omega & 0 & 0 & \Omega & \Omega & 0 & \Omega & \Omega & \Delta & \Delta & \varepsilon^2_{-\vec{Q}_1} & 0 & \Delta & \Delta & 0 & \Delta & \Delta \\
\Omega & \Omega & 0 & 0 & \Omega & \Omega & 0 & \Omega & \Omega & \Delta & \Delta & 0 & \varepsilon^2_{\vec{Q}_2} & \Delta & \Delta & 0 & \Delta & \Delta \\
\Omega & 0 & \Omega & \Omega & 0 & 0 & \Omega & \Omega & \Omega & \Delta & 0 & \Delta & \Delta & \varepsilon^2_{-\vec{Q}_2} & 0 & \Delta & \Delta & \Delta \\
\Omega & 0 & \Omega & \Omega & 0 & 0 & \Omega & \Omega & \Omega & \Delta & 0 & \Delta & \Delta & 0 & \varepsilon^2_{\vec{Q}_3} & \Delta & \Delta & \Delta \\
\Omega & \Omega & 0 & 0 & \Omega & \Omega & 0 & \Omega & \Omega & \Delta & \Delta & 0 & 0 & \Delta & \Delta & \varepsilon^2_{-\vec{Q}_3} & \Delta & \Delta \\
0 & \Omega & \Omega & \Omega & \Omega & \Omega & \Omega & 0 & 0 & 0 & \Delta & \Delta & \Delta & \Delta & \Delta & \Delta & \varepsilon^2_{\vec{Q}_4} & 0 \\
0 & \Omega & \Omega & \Omega & \Omega & \Omega & \Omega & 0 & 0 & 0 & \Delta & \Delta & \Delta & \Delta & \Delta & \Delta & 0 & \varepsilon^2_{-\vec{Q}_4}
\end{pmatrix} \quad \text{(S5)}$$

Here, $\varepsilon^a{}_b$ is shorthand for $\varepsilon^a(\vec{k}\text{-}\vec{b})$ for formatting reasons. $a$ = 1,2 refers to the band in the ARPES tight binding model and $\vec{k}\text{-}\vec{b}$ refers to the *k*-position where the dispersion is evaluated, effectively shifting the dispersion by $+\vec{b}$. Using Δ = Ω = 3 meV the resulting FS and new BZ are shown in Fig. S10**A** below. An overview of the properties of the various pockets before and after reconstruction are listed in Table S1.

| $f$ (T) | mass ($m_0$) | LK damping | type | Carrier density $n$ ($10^{26}$ m$^{-3}$) | # copies | location | $B_{QO,l}$ (T) | $B_{QO,\tau}$ (T) | $R_c$ at 30 T |
|---|---|---|---|---|---|---|---|---|---|
| 3907 | 2.49 | 0.978 | hole | 15.1 | 2 | 1 $K$ | | 30 | 75 |
| 7490 | 2.28 | 0.981 | hole | 29.0 | 1 | 1 $\Gamma$ | | 58 | 104 |
| 7604 | 2.42 | 0.979 | hole | 29.4 | 2 | 2 $K$ | | 45 | 105 |
| 9522 | 2.74 | 0.973 | hole | 36.9 | 1 | 2 $\Gamma$ | | 69 | 117 |
| 48 | 0.30 | 1.000 | hole | 0.185 | 6 | $K^*$ | 4.8 | 5.8 | 8.3 |
| 315 | 0.71 | 0.998 | hole | 1.215 | 2 | $K^*$ | 12.5 | 14.3 | 21 |
| 639 | 1.51 | 0.992 | hole | 2.462 | 2 | $K^*$ | 25.4 | 29.6 | 30 |
| 41 | 0.68 | 0.998 | electron | 0.157 | 3 | $M^*$ | 8.5 | 13.5 | 7.7 |
| 1079 | 0.86 | 0.997 | electron | 4.161 | 1 | $\Gamma$ | 21.2 | 16.8 | 40 |
| 1706 | 1.26 | 0.994 | electron | 6.575 | 1 | $\Gamma$ | 28.1 | 24.3 | 50 |
| 1907 | 1.24 | 0.994 | electron | 7.350 | 1 | $\Gamma$ | 28.3 | 23.8 | 53 |
| 1957 | 1.28 | 0.994 | electron | 7.545 | 1 | $\Gamma$ | 29.0 | 25.0 | 53 |

**Table S1**. Quantitative information per pocket. The top section refers to the unreconstructed FS (*35*), the bottom section to the reconstructed FS shown in Fig. S10A. Here, *c*-axis warping and the pancake pocket are neglected. The QO frequency $f$ assumes $\vec{H}//c$ determined via Onsager's relation $f = \hbar A_k/2\pi e$ where $A_k$ is the cross-sectional area of the pocket. The effective cyclotron mass is determined via $1/m^* = \frac{\hbar^2}{2\pi}\frac{\mathrm{d}A_k}{\mathrm{d}\varepsilon}$ at the Fermi level. $m_0$ is the vacuum electron mass. LK stands for Lifshitz-Kosevich and this column shows the size of the QOs compared to zero temperature expected from damping at 30 T and 0.3 K. We conclude that thermal damping is not the reason why QOs are not observed. Carrier type is determined by whether the cross section of the orbit grows or shrinks with increasing energy. The carrier density is listed per copy but includes spin degeneracy via $A_k/2\pi^2 c$. The location column lists in the unreconstructed case the band number from the tight binding model (*35*) and the high symmetry point around or nearest to which the orbit is located. Stars (*) refer to the high symmetry location in the reconstructed BZ (where $\Gamma^* = \Gamma$). $B_{QO}$ is the onset magnetic field where QOs are expected. We define $\omega_c\tau = 2\pi$ as the magnetic field at which charge has a probability $1/e \sim 0.37$ to complete an orbit. QOs are expected when $\omega_c\tau = 1$ while matching $\rho_{ab} = 2.5$ $\mu\Omega$cm at 0 T is estimated as the residual resistivity. The first estimate assumes the unimpeded isotropic-$\ell$ scenario of Fig S10 ($\ell$ = 60 nm), the second estimate assumes the unimpeded isotropic-$\tau$ scenario of Fig. S11 $\tau$ =3.0 ps). Note that experimentally we exceed both estimates of $B_{QO}$ for all eight reconstructed Fermi pockets yet observe no QOs. $R_c = \hbar\bar{k}/Be$ is the average real-space cyclotron radius, where $\bar{k} = \sqrt{A_k/\pi}$ approximates the cross section of the orbit by a circle (independent of scattering model).

| | 10 K | 15 K | 20 K | 25 K | 30 K | 10 K (Fig. 3) | 25 K (Fig. 3) |
|---|---|---|---|---|---|---|---|
| $\tau_c$ (ps) | 0.223 | 0.170 | 0.122 | 0.130 | 0.133 | 0.198 | 0.0575 |
| $\tau_h$ (ps) | ≤0.0002 | ~0.00035 | 0.0015 | 0.01 | 0.02 | ≤0.0002 | 0.00689 |
| $\epsilon_\Delta$ (meV) | ~5 | ~5 | 10 | 50 | 100 | 20 | 20 |
| $\rho_{\text{model}}(0)$ ($\mu\Omega\text{cm}$) | 2.66 | 3.49 | 5.09 | 6.94 | 8.89 | 3.88 | 11.83 |
| $\rho^*_{\text{exp}}(0)$ ($\mu\Omega\text{cm}$) | 2.67 | 3.52 | 5.09 | 6.93 | 8.88 | 3.90 | 11.83 |

**Table S2:** Parameters used in Fig. S8 (to fit data for the *RRR* = 41 sample). The model is entirely defined by $\tau_c$, $\tau_h$ and $\epsilon_\Delta$. $\epsilon_\Delta$ controls the hotspot width and is used to fit the low-field $H^2$ coefficient of the MR (see Fig. S7); $\tau_h$ is used to fit the breakdown field where $d\rho/d\mu_0 H$ is at a maximum, and $\tau_c$ is fitted to $\rho(0)$ as listed. None of these parameters can be used to meaningfully modify the LMR slope itself. At the lowest temperatures, no breakdown is observed and the hotspots themselves are not fitted. Instead, a guess is taken for $\epsilon_\Delta$ based on the high-*T* data. $\tau_h$ is subsequently sufficiently small that no breakdown occurs in the accessible field range. As stated in the main text, at low *T* the model has only one meaningful degree of freedom, namely $\tau_c$. Taking $\langle v\rangle \approx 2\times 10^5$ m/s, the Mott-Ioffe-Regel limit may be defined as the lifetime (≈ 0.0015 ps) where the mean-free-path equals $a$, the unit cell parameter. The temperatures with meaningful data relating to $\tau_h$ satisfy this limit. The lowest *T* results can be tuned above this limit by broadening the hot spots. The last two columns give the parameters used for the fits in Fig. 3 of the main article.

Next, we compute the conductivity of this reconstructed FS. The simplest possible analysis is to neglect all the distorted shapes and apply Drude theory, treating each pocket as a perfect circle. This approximation appears crude in light of the pronounced anisotropy in the shapes shown in Fig. S10**A**. However, we will show with detailed calculations below that this simplistic calculation nevertheless captures a number of properties of the magnetotransport that are tied directly to the semimetallic nature and to which the FS anisotropy is largely irrelevant. Thus, this simplistic first analysis serves to understand which aspects of the model result from semimetallicity and which from FS anisotropy. The conductivity contribution of each pocket is then simply $\sigma_{xx} = \sigma_{yy} = ne^2\tau/m^*/(1+\omega_c^2\tau^2)$ and $\sigma_{xy} = -\sigma_{yx} = \omega_c\tau\sigma_{xx}$. For each pocket, we use the number of copies in the BZ as well as carrier density $n$, carrier type (sign of $\omega_c\tau$) and effective mass $m^*$ listed in Table S1. $\omega_c = qB/m^*$ is the cyclotron frequency where $q = \pm e$ is signed. The only degree of freedom is an isotropic $\tau$ = 0.36 ps for all pockets, consistent with a realistic zero-field resistivity of 3.0 $\mu\Omega$cm at 10K. This lifetime is comparable to the value used without reconstruction due to the low fraction of the density of states gapped by the CDW order. After matrix inversion, the resulting magnetotransport is shown in Fig. S9.

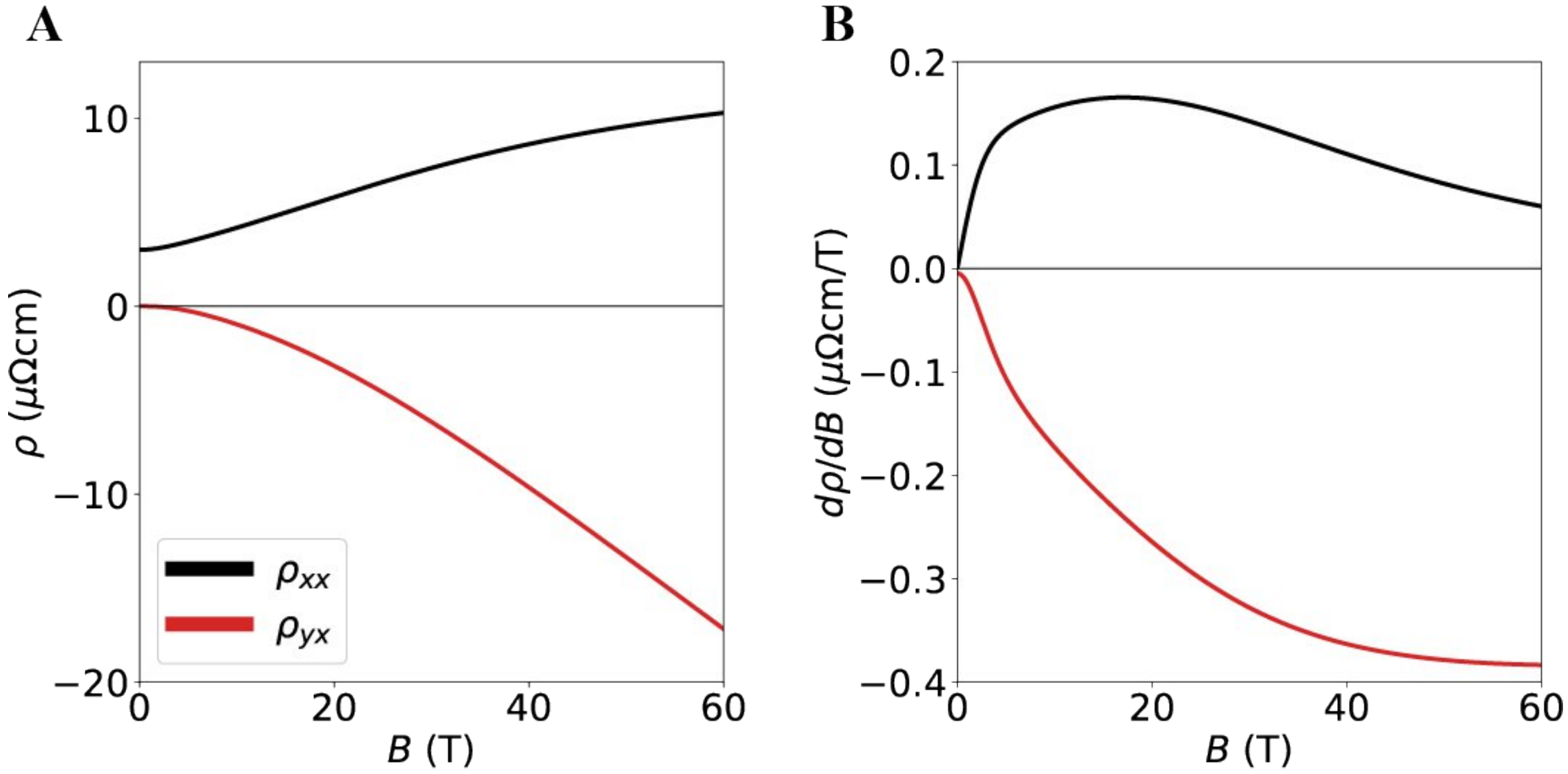

**Fig. S9**. MR and Hall resistivity post-reconstruction using Drude theory. A) $\rho_{xx}(H)$ and $\rho_{yx}(H)$ using the carrier density, effective mass and number of copies of each pocket in the BZ (see Table S1). The only degree of freedom is the (isotropic) scattering lifetime $\tau$ = 0.36 ps corresponding to $\rho_{ab}(0)$ = 3 $\mu\Omega$cm. B) Corresponding derivatives. $R_H$ can be obtained in mm$^3$/C by multiplying the red curve by 10. Notice that the Hall coefficient is negative yet vanishing at low field and double the size shown in the data in Fig. S1 at high fields. Meanwhile, the slope of the MR is of the right size, yet no extended *H*-linear regime exists and the quadratic MR at lowest magnetic field extends to over 5 T.

We point out a number of aspects associated with this result. Firstly, the Hall effect is negative as a result of the 3:1 electron:hole ratio. Secondly, the size of the negative Hall effect deviates from experiment as the theoretical result vanishes at low magnetic field and is double the experimental value at high magnetic fields indicating fewer holes are actually present at the Fermi level. Thirdly, the MR is of the correct order of magnitude but saturates and the quadratic regime extends to a field strength that is too high. The tendency of the semimetallic MR to grow large despite only a small admixture of holes is well known (*13*). Finally, there is no extended, non-saturating *H*-linear regime in the MR curve.

Next, we perform detailed resistivity calculations on the reconstructed FS. We employ the full SCTIF formalism of Eq. (S1) as above for impeded cyclotron motion. However, some of the reconstructed pockets are sufficiently anomalously shaped that they backtrack in $\varphi$ during their orbits relative to any origin placed inside. A single $\varphi$ value may thus refer to multiple points on the orbit and at inflection points any $\varphi$-derivative diverges as a result of this loss of bijectivity. We therefore embrace the fully *k*-space path-length approach without any underlying parameterization using azimuthal angle $\varphi$ for all integrals. We also rely solely on the mean free path and FS shape. No reference to velocity, $\omega_c$, $\tau$ or $\omega_c\tau$ is required. We achieve both aims by expanding the Ong formalism (*66*) beyond the low-field limit. Without approximation, we substitute time in favor of path- length in *k*-space along the orbit $k_o$ using the equations of motion in Eq. (S1). We use the fact that velocity and magnetic field are always orthogonal in the present context to simplify this step $\hbar\dot{k}_o = qvB$, where subscript *o* refers to the path-length along the orbit towards the past:

$$\sigma_{ij} = \frac{e^2}{\pi\hbar}\int_{\mathrm{FS}} \frac{\mathrm{d}^2\vec{k}}{(2\pi)^2}\frac{\ell_i}{\ell}\int_0^\infty \mathrm{d}k_o\hbar\frac{\ell_j(k_o)}{eB\ell(k_0)}\ \exp\left(-\int_0^{k_o}\frac{\hbar dk_o'}{eB\ell(k_0')}\right) \tag{S6}$$

Here, $\ell$ is the size of the mean-free-path and $\ell_i$ the $x$ or $y$ component. This formula allows us to compute the conductivity using solely the FS shape and mean-free-path vector across it (whose direction is set perpendicular to the local curvature). By using solely the information that is absolutely required, we are able to keep the number of degrees of freedom to an absolute minimum. On top of this, we again employ time-ordered custom Simpson integration but now in terms of $k_o$ to enable efficient integration without assumption about the magnetic field strength. Although $B = 0$ is strictly ill-defined, the limit $B \to 0$ is exactly the same as the SCTIF and in practice we are able to maintain numerical accuracy down to $10^{-5}$ T.

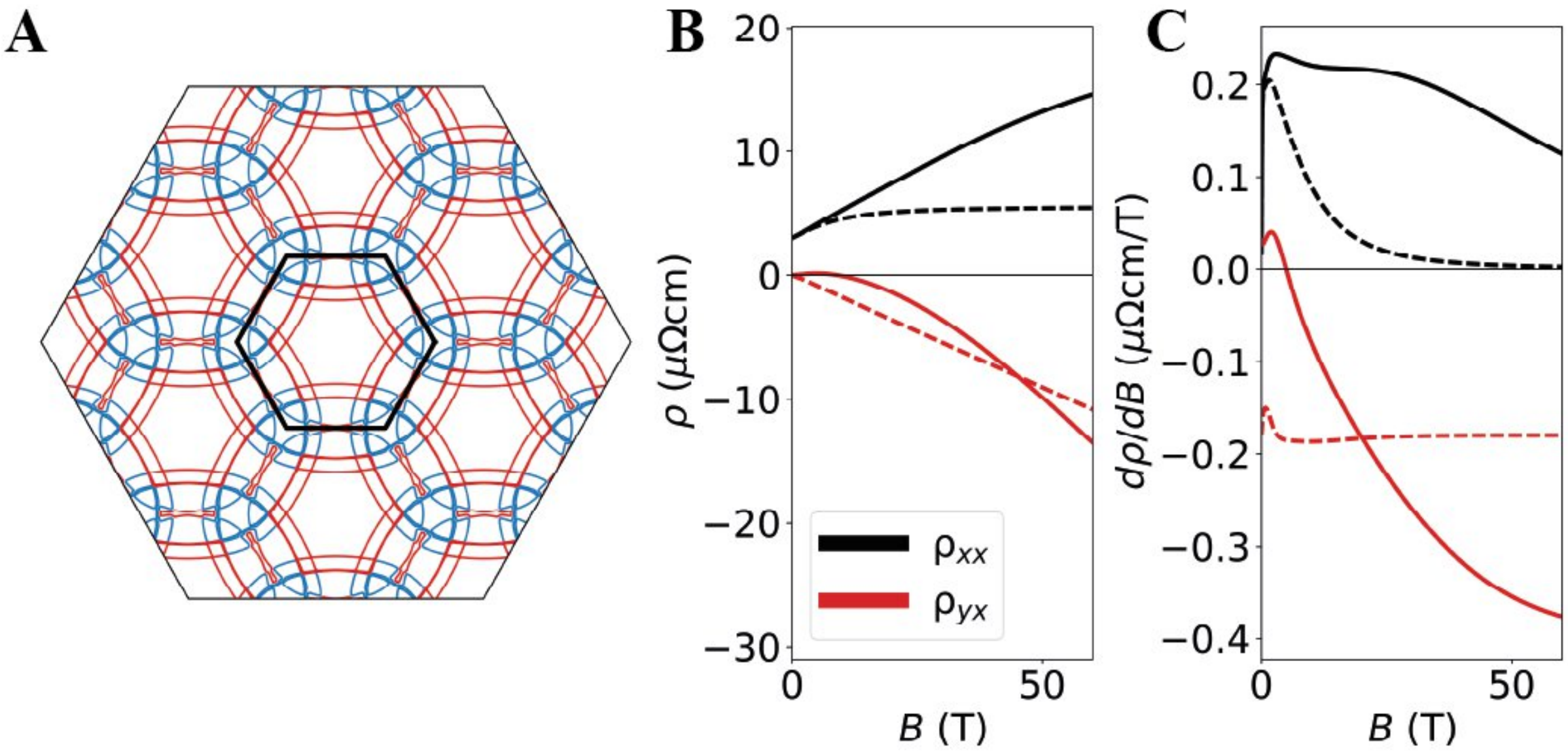

**Fig. S10**. Peierls scenario under isotropic-$\ell$. A) Fermi surface after reconstruction. The outer hexagon is the normal-state BZ, the small hexagon the BZ in the CDW state neglecting the incommensurability. Blue lines represent hole-like pockets, red lines electron-like pockets. B) MR and Hall resistivity after matrix inversion obtained from Eq. (S6) using isotropic $\ell$ = 50 nm. The mean-free-path is the only degree of freedom and chosen to obtain $\rho_{ab}(0)$ = 3 $\mu\Omega$cm. Dashed lines represents the same MR calculation, but with all pockets rendered artificially electron-like, showing the large influence of the semi-metallic nature, especially at high field. C) Corresponding derivatives. Compared to Fig. S9 the low-field quadratic MR is strongly enhanced due to anisotropy of the FS, while the high-field MR is almost identical.

To fully define the model, we employ an isotropic $\ell$ across the reconstructed FS. Because the magnetotransport is ultimately determined by the shape of the FS and the mean-free-path vector across it, comparing an isotropic $\ell$ to the Drude model essentially highlights the impact of the complex Fermi pocket shape formed by the reconstruction. An isotropic $\ell$ is the nominal expectation for the scattering rate at low $T$ where only elastic scattering off point-like impurities is relevant. The result is shown in Fig. S10.

Comparing the results with the Drude model (Fig. S9), we find that the high-field regime above about 15 T is similar. This is expected as high $\omega_c\tau$ means that quasiparticles traverse most of the Fermi pocket before decaying, washing out the influence of anisotropy. While the high-field regime is similar to the Drude model, the anisotropy of the FS leaves a pronounced mark on the low-field MR. We find a steep quadratic MR up to ~0.4 T compared to the much more gradual quadratic MR up to ~7 T in the Drude scenario. This strong modification of the low-field MR thus emerges as the leading candidate to explain the low turnover scale $H^*$ found experimentally compared to the simple theoretical impeded cyclotron orbital model in the main article.

To further differentiate the impact of the FS anisotropy versus the electron-hole mixture, we change all pockets to electron-like and show the resulting magnetotransport in the dashed lines of Fig. S10. No other changes were made. The anisotropy of the FS shape indeed determines the entire MR below 3 T, yet without semimetallicity the MR would quickly saturate above 5 T. Thus, the sharp corners in the FS cannot explain the non-saturating $H$-linear MR up to 30 T observed in experiment. Indeed, Pippard's classic work on magnetotransport clearly outlines that the linear MR from sharp FS corners saturates when quasiparticles can traverse between corners, i.e. when $\omega_c\tau$ approaches the corner-to-corner distance. This saturation field for FS corners is therefore considerably lower than the magnetic field where QOs appear, which requires complete cyclotron orbits. The origin of the extended region of quasi-linearity in Fig. S10 compared to the Drude model (Fig. S9) thus arises from two separate peaks: A 1-5 T MR peak due to FS anisotropy, and a 7-20 T peak due to the semimetallic nature of the FS. There is no *a priori* reason why these two quasi-linear slopes need to be the same and some deviation from $H$-linearity in the transition region would thus be expected in $d\rho_{ab}/d\mu_0H$. However, experimentally the $H$-linear slope appears entirely constant as a function of magnetic field, even well above 20 T using derivatives. We point out that the vanishing Hall coefficient at low magnetic fields and its large value at high magnetic fields both suggest the theoretical model overestimates the number of holes substantially, which would reduce the semimetallic MR above 5 T.

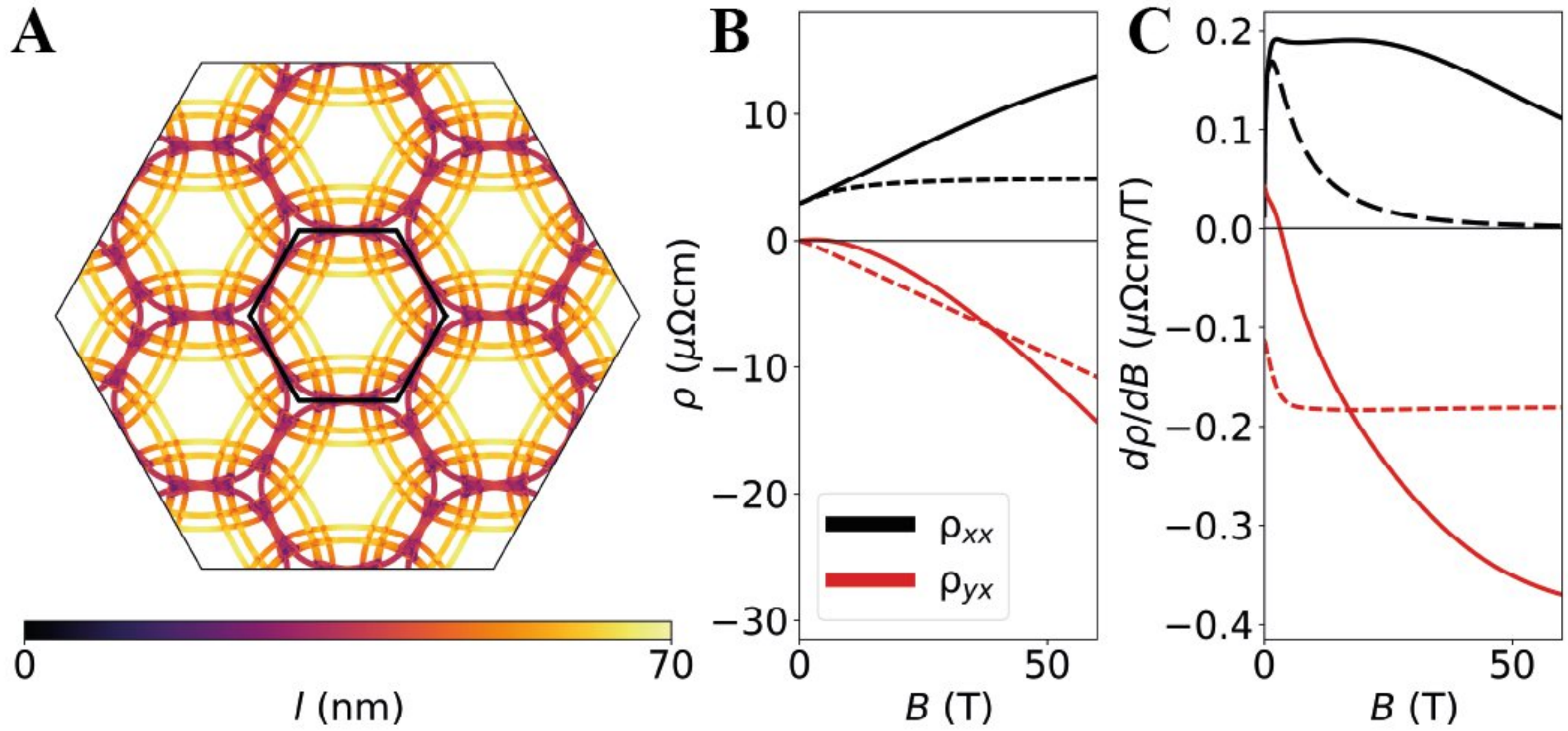

**Fig. S11**. Peierls scenario with isotropic $\tau$. A) FS with the mean-free-path (Fermi velocity anisotropy) displayed in color. This anisotropy is obtained using $l = v_F\tau$ where isotropic $\tau = 0.25$ ps is the only degree of freedom chosen to match $\rho_{ab}(0) = 3$ μΩcm. B) Resistivity after matrix inversion obtained from Eq. (S6). Dashed lines represent the same MR calculation, but all pockets are artificially electron-like, showing the large influence of the semi-metallic nature, especially at high field. C) Corresponding derivatives. We obtain largely the same magnetotransport as shown in Fig S9.

We further evaluate this model for the case of isotropic $\tau$ using the Fermi velocities of the ARPES-derived dispersion relation to obtain the mean-free-path (*35*). This additional evaluation tests the robustness of these conclusions under anisotropy in $\ell$. In particular, where the FS hybridizes we find up to 2× suppression of the Fermi velocity, which may lead to a locally reduced $\ell$. The variation of $\ell$ around the FS in this case as well as the magnetotransport results are shown in Fig. S11. Except for details at the lowest magnetic field due to a different mobility balance between the pockets, we observe almost identical results to the isotropic-$\ell$ model. This result shows that the suppression of $v_F$ which naturally emerges where the FS reconstructs in the Peierls picture is by itself insufficient to create effective impedances to orbital motion. Above a few Tesla, the anisotropy of the FS quickly washes out, the MR saturates and QOs are expected.

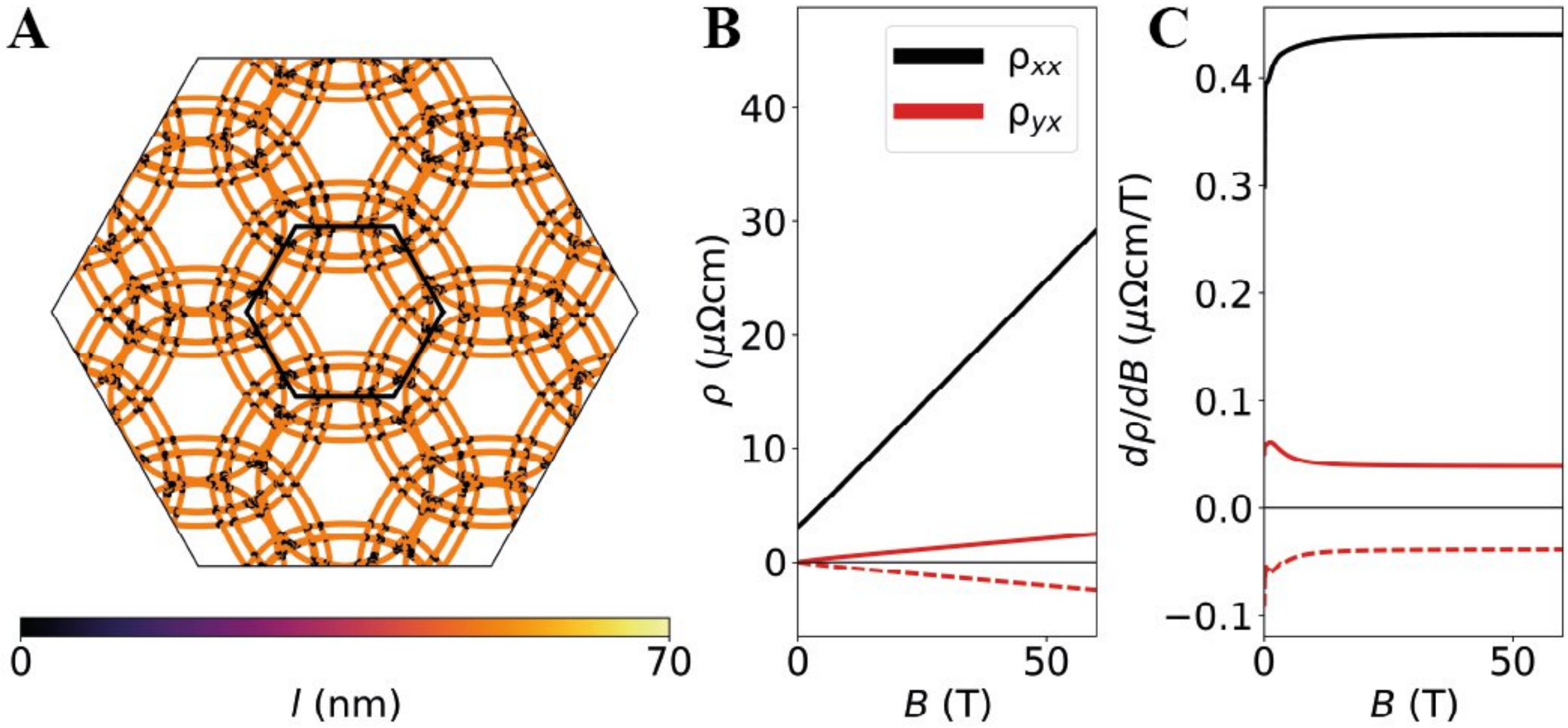

**Fig. S12**. Peierls scenario with impedances. A) FS with the mean-free-path displayed in color. The model is the same as in Fig. S10, except that where the FS hybridizes impedances are added. The zero-field resistivity rises by 3% due to the hotspots. B) Resistivity after matrix inversion obtained from Eq. (S1). Dashed line represents the same MR calculation, but with all pockets rendered artificially electron-like. Unlike above, the semi-metallicity is of minor importance to the MR and the Hall effect is hole-like, showing that impeded cyclotron motion dominates the magnetotransport. The semimetallic nature is largely irrelevant because $\omega_c\tau$ cannot grow larger than the inter-hotspot distance. C) Corresponding derivative. Unlike in Fig. S10, the MR is robustly linear to highest magnetic field.

The fourth and final iteration of this model is to implement effective impedances at the points where the FS reconstructs. Due to the complexity of the FS reconstruction, the only coherent way we found to approach this problem is by introducing hotspots wherever the FS folds back onto itself, including between pockets and bands. Impedances are again given a Gaussian shape as above, but with a fixed $k$-radius of $2\times10^6$ $m^{-1}$ instead of an energy criterion which is no longer accessible. This results in many more impedances than shown in the main article, where only intrapocket hotspots are considered. The result is robust LMR as shown in Fig. S12 albeit with a slope that is 3 times too high. The Hall effect in this impeded version is back to hole-like despite the full FS reconstruction due to the effective suppression of turning points on the FS which dominate the Hall response. We thus suggest that the Peierls reconstruction takes place, but that effective impedances are only formed where ARPES shows a clear CDW gap. This led us to the model shown in the main article. At this time, unifying impeded cyclotron motion and the Peierls reconstruction is not possible phenomenologically and likely requires insight into the microscopic interactions which drive both phenomena.

In this section we have presented a detailed evaluation of the expected consequences of the Peierls reconstruction on the magnetotransport and QOs in $NbSe_2$. These calculations show that sharp FS corners and semimetallicity do leave a substantial mark on the MR, but that the results are not in agreement with experiment. In particular, without impedances, we expect the MR to change $H$-linear slope around 7 T, saturate above 20 T and for a myriad of QOs to appear below 30 T. These aspects are all incompatible with the experimental observations.

C. Beyond the RTA

The success of this simple impeded cyclotron motion model within the RTA to explain the observed MR and lack of QOs is noteworthy. With only the cold scattering rate as a single fit parameter it is able to account for many aspects of the experimental data, including the *H*-linear MR, the low turnover scale, the Kohler scaling, the low- and high- temperature MR, the qualitative behavior at the CDW transition and the lack of QOs in clean samples. This consistency with a wide range of experimental results with minimal degrees of freedom is our main result.

Nevertheless, the model remains phenomenological and has limitations. In this section we dive deeper into the following physical aspects; the non-collinear velocity between the endpoints of hotspots and the possibility of small-angle scattering particularly at intermediate temperatures (both explained below). In order to progress our understanding of the MR in $NbSe_2$ as much as possible, we here present additional state-of-the-art modeling beyond the RTA to explore these open questions. We emphasize, however, that in our experience thus far these more complex models fail to account for additional features in the experiment. Nevertheless, these computations serve to imply that small-angle scattering is not relevant to $NbSe_2$ and that the removal of spectral weight or another non-scattering origin is more likely than scattering off fluctuations as the source of the impedances to cyclotron motion.

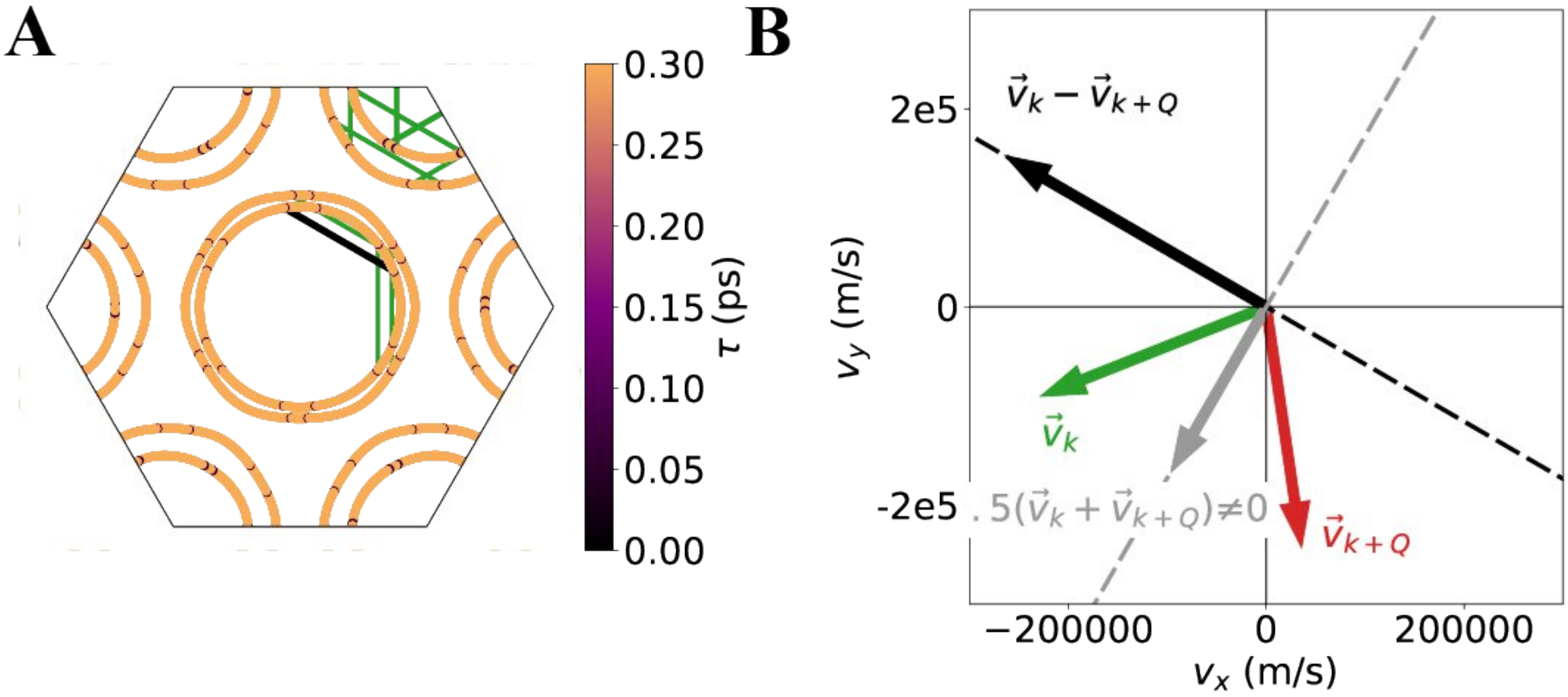

**Fig. S13**. Non-collinear velocity. A) Similar to Fig. 1B of the main article with one hotspot pair marked by the black line. The bottom right end of the black line is referred to as $\vec{v}$ and top left is $\vec{v}+\vec{Q}$. B) The vector sum and difference of the corresponding velocities. At the hotspot, the black difference vector decays fast and the gray sum vector decays slowly. The latter is presumed to vanish in any RTA model. Consequently, a quasiparticle starting out as either green or red will have their net velocity direction quickly rotate to the gray direction and then slowly decay, resulting in a net current contribution which is no longer along the original velocity direction, i.e. $\vec{l}$ cannot be written as $\vec{v}\tau$ for any value of $\tau$ in violation of the RTA.

One clear theoretical limitation of the RTA is non-parallel velocity between the endpoints of hotspots; i.e. $\vec{v}_{\vec{k}}+\vec{v}_{\vec{k}+\vec{Q}}\neq 0$. This is visualized in Fig. S13. Consider the scenario where the quasiparticle is quickly scattered back and forth between the two endpoints of a specific hotspot due to e.g. fluctuations or domain wall scattering. After a short time, this results in an even probability for the quasiparticle to maintain $\vec{v}_{\vec{k}}$ or $\vec{v}_{\vec{k}+\vec{Q}}$. In the RTA, the assumption is made that after this time no net contribution to the current remains, i.e. the average velocity vanishes or $\vec{v}_{\vec{k}}+\vec{v}_{\vec{k}+\vec{Q}} = 0$. In quasi-1D systems this assumption is reasonable given the predominantly anti-parallel and similarly sized velocities on the right-moving and left-moving FS sheets. However, $NbSe_2$ is a quasi-2D material, however, and as such, strong $\vec{v}_{\vec{k}} + \vec{v}_{\vec{k}+\vec{Q}} \neq 0$ violations naturally occur. For the highlighted hotspot in Fig. S13, the velocity component indeed relaxes according to the fast or hot scattering rate along the black dashed direction, but a substantial component $\vec{v}_{\vec{k}} + \vec{v}_{\vec{k}+\vec{Q}} \neq 0$ remains in the direction of the gray dashed line which instead decays according to the much slower background cold scattering rate. The non-collinear velocity thus results in a rotation rather than complete decay of the net velocity and $\vec{l}$ and $\vec{v}$ are no longer aligned as assumed in the RTA. We thus obtain a bifurcation of the relaxation time for different velocity directions at fixed $\vec{k}$, which is beyond the (standard) RTA and few works exist which explore this regime.

This raises an important question about the nature of the hotspots: If the impedances are caused by a shift of spectral weight away from the Fermi level or otherwise terminate quasiparticle coherence, then it is appropriately modeled by impedances in the relaxation time as presented in the main article. However, if the impedances merely scatter charge back and forth elastically between two hotspots akin to a fluctuation regime usually found above $T_{CDW}$, then the non-collinear velocity at hotspots will result in further contributions to the conductivity. In Ref. (*23*,*47*,*67*) it was shown that exactly such non-collinear velocities at hotspots result in large corrections to the Hall effect and may even explain the sign change in electron-doped cuprates. The calculations in the RTA have implicitly assumed the first scenario by effectively impeding the charge at the impedance. Is impeded cyclotron motion robust under these latter effects?

In this section, we introduce a new Boltzmann framework to investigate the consequences of the latter hypothesis (scattering hotspot). This framework generalizes previous work aimed at investigating small-angle scattering, which we therefore also consider. We show that 1) scattering hotspots remain effective impedances to cyclotron motion but with increased turnover scale and reduced LMR, 2) the resulting MR is inconsistent with experiment, 3) that small-angle scattering maintains effective impedances with the same LMR but with a higher turnover scale and 4) small-angle scattering matches experiment less well than large-angle scattering.

D. Developing the Boltzmann transport framework

We have outlined above that the main issue to be investigated is the non-collinear velocity at hotspots. In Ref. (*67*) this was investigated using Kubo formalism in the low-field limit. Here, $\omega_c\tau$ is sufficiently high however that the problem is more difficult. After colliding with a hotspot, charge will evolve under cyclotron motion and a substantial probability remains for charge to reach another hotspot and cascade into a complex pattern. However, the underlying physics can be modeled semiclassically, which we develop and use to move past the low-field limit.

We develop a computational implementation of old ideas by Pippard, who investigated the effect of small-angle scattering on the MR of elemental Cu (*68*). The central idea is to abandon an absolute position of a quasiparticle on the FS, but instead to define a probability distribution across the FS where the quasiparticle resides. This probability distribution starts as a delta-function and evolves over time in the statistical average of ~$10^{23}$ electrons. Under small-angle scattering the appropriate process is diffusion; for hotspots, we simply include scattering matrix elements beyond nearest neighbors of the appropriate magnitude. The framework is strictly a generalization of the SCTIF (*45*,*46*). This new framework was recently developed by the present authors to investigate intraband (small-angle) vs interband (Umklapp) scattering in quasi-1D conductors, which results in a bifurcation of the RTA (*48*). Here, we expand the framework from quasi-1D to 2D and beyond the low-field ($\omega_c\tau \leq 0.1$) regime.

To proceed, we define a probability distribution across the FS, which is taken to be 2D and therefore a line. No objection exists to define a surface-distribution for a 3D FS other than computational complexity.

$$\int_{FS} \mathrm{d}k P(k) = 1 \tag{S7}$$

where $P(k) \geq 0$ at all times. As in the RTA, we use the FS from Ref. (*35*). This probability distribution is not a superposition since phase information is neglected. $P(k) = |\Psi(k)|^2$ is more accurate since both represent quasiparticle probabilities, but we emphasize that $P$ is considered in the statistical average of all electrons in the metal. For instance, elastic small-angle scattering from impurities results in Brownian motion of electrons across the FS (*68*). In its extreme small-angle limit, this process is modeled through diffusion of the probability distribution $P$ across the FS, starting as a delta-function at zero time and turning into a homogeneous distribution at late times. This contrasts with the RTA in the SCTIF, in which charge is considered to no longer contribute at all to the conductivity after a scattering event and no preferential or structured scattering is considered. A key difference is also that in SCTIF at late times no charge remains, whereas in the new framework charge is strictly conserved and at late times the ensemble average velocity simply vanishes as $P(k)$ approaches thermal equilibrium. Eq. (S1) is rewritten as

$$\sigma_{ij} = \frac{e^2}{\pi\hbar}\int_{\mathrm{FS}} \frac{\mathrm{d}^2\vec{k}}{(2\pi)^2}\frac{v_i}{v}\int_0^\infty \mathrm{d}t\int_{FS}\mathrm{d}k' v_j(k') P(k',t) \tag{S8}$$

SCTIF is formally recovered by using $P(k',t) = \delta(k'-k(-t))\exp(\int_0^t \mathrm{d}t'/\tau(-t'))$. Thus, the norm of $P$ (the total charge) exponentially vanishes as the charge is unaccounted for after scattering in the RTA. However, this removal of charge from the framework is no longer necessary. We may instead maintain strict charge conservation $Q = \int_{FS}\mathrm{d}k P(k) = 1$ at all times and diffuse the quasiparticle probability across the FS such that at late times $\langle\vec{v}\rangle = 0$, where the average represents integration of the probability over the FS $\langle\vec{v}\rangle = \int_{FS}\mathrm{d}k P(k)\,\vec{v}(k)$. This formalism neglects the phase of the wave function as well as any energetics of the dynamics. These two simplifications are linked, by neglecting the energetics of any dynamics, no phase differences can accrue from $\exp(-i\Delta\varepsilon t/\hbar)$ phases due to a quasiparticle spending a time $t$ at energy $\Delta\varepsilon$ away from the Fermi level. In this manner, a wide variety of charge-conserving dynamics can be implemented in a semiclassical fashion beyond the RTA. This includes the small-angle scattering and hotspot scattering of interest here.

Using this extension of the SCTIF, we incorporate scattering dynamics beyond the RTA on $P(\vec{v},t)$.

$$\frac{\partial P}{\partial t} = D\frac{\partial^2 P}{\partial k^2} + \frac{1}{\tau_h}\sum_i \left(P\left(\vec{k}+\vec{Q}_i\right) - P\right)\delta\left(\varepsilon_{\vec{k}+\vec{Q}_i}\right) \tag{S9}$$

The first term is diffusion with coefficient $D$ setting the cold scattering rate, the second term is hotspot scattering, which is only active where a $Q$-vector connects the FS. Where not shown, the time and $k$ dependence of $P$ are $P\left(\vec{k},t\right)$ for clarity. The RTA of both terms is implemented as follows

$$\frac{\partial P}{\partial t} = -\frac{P}{\tau_c} - \frac{1}{\tau_h}\sum_i P\delta\left(\varepsilon_{\vec{k}+\vec{Q}_i}\right) \tag{S10}$$

In both approaches, the scattering matrix contains two terms. The first term is the cold scattering rate. In the RTA, this follows the exponential decay equivalent to large-angle scattering, while beyond the RTA, we follow Pippard (*68*) and implement small-angle scattering in its extreme limit, namely as a diffusion of the probability $P(k)$ over time. Large-angle scattering can be implemented as well in the new formulism, e.g. isotropic using $\frac{1}{\tau}\sum_i\left(P\left(k_i\right) - P\right)$. The second term is the hot scattering rate which only acts where the FS is connected by one of the $Q$-vectors. In the RTA this also follows exponential quasiparticle decay, while beyond the RTA we preserve charge and scatter between the endpoints of the hotspot. All scattering terms are implemented by discretizing the FS and creating a scattering matrix $\dot{P}_i = M_{ij}P_j$ of the momentary derivative. We then use Runge-Kutta 4 to evolve the probability distribution by a discrete timestep. The fundamental timesteps taken are typically 1/100 of the hot scattering time. In order to progress to the time where the velocity has fully relaxed (about 10× the cold scattering time), we must raise $M$ to a power of over $10^9$, which is achieved in about $10^4$ (unequally spaced) steps in order to integrate over time. Between steps we evaluate up to $M^{10^6}$ using binary exponentiation to save time on raising powers of the matrices.

Next, we implement cyclotron motion. In Ref. (*48*) both cyclotron motion and scattering were incorporated through the equations of motion. Over a given timestep, however, the discrete $k$-shift due to cyclotron motion and the *k*-distance between subsequent points on the discretized FS will not coincide. Nevertheless, in order to obtain the desired net rotation of the quasiparticle distribution *P*, one must split the probability into two proportions. For example, if the distance between two *k*-states is $1\times10^{7}$ $m^{-1}$ yet the cyclotron motion over an individual timestep is $10^{6}$ $m^{-1}$, then 10% of the total probability is transferred in the direction of cyclotron motion to the nearest neighbor. In the absence of any scattering, however, this partial transfer will result over time in a broadening of the probability distribution and introduce an artificial diffusion process limiting quasiparticles to about one rotation around the FS independent of the number of FS points; $\omega_c\tau \leq 2\pi$ (this artificial decay of $\langle v\rangle$ follows an exponential due to the central limit theorem). This artifact was not an issue in Ref. (*48*) since the primary results required $\omega_c\tau < 0.1$. For impeded cyclotron motion in $NbSe_2$, however, this is insufficient.

Here, we solve this problem and expand the validity of the model to arbitrary magnetic field strength. Ultimately, cyclotron motion is not a scattering or probabilistic event but rather a set rotation of all states. This is achieved by shifting the underlying basis, the discrete points on the FS themselves, at every time step. This change is non-trivial because the distance between subsequent *k*-positions shrinks or expands under cyclotron motion with the anisotropy of the effective mass. We therefore must renormalize the weights $P_i$ such that the total probability for charge to be in a specific interval remains constant under cyclotron motion during each step and no charge is created or removed. At the same time, the movement of *k*-states requires re-computing the scattering matrix *M* during each evaluation, significantly increasing the computational cost. Finally, diffusion processes must be implemented in a stable fashion, which takes extra consideration due to the necessarily unequal distribution of the *k* points on the FS which is additionally changing as a function of time due to the small but finite anisotropy of the FS from Ref. (*35*). Details and the full implementation can be found in Ref. (*65*). The code is tested against Drude theory (including under magnetic field for Hall effect and MR) as well as impeded cyclotron motion and reproduces the results of the main article and Ref. (*21*) if Eq. (S10) is used to reproduce the RTA results.

In Fig. S14 we show the resulting MR in four different regimes. The first regime reproduces the RTA results of the main article within the new formalism. In this scenario, charge decays exponentially for both cold and hot scattering, albeit with different timescales. The result is robust LMR. Next, we introduce small-angle scattering for the cold charge. Physically, this is relevant above the temperature where phonon scattering dominates over impurity scattering, while remaining well below the Debye temperature where phonons of sufficiently large *q* are thermally available to scatter all across the FS. Small-angle scattering is also relevant if the impurities are believed to reside between the 2D layers, as in cuprates (*69*). The result is again LMR of similar magnitude due to the continued existence of the hotspots, but with a prolonged quadratic regime at low field, in contrast with the experimental observation. We also note that with small-angle scattering the hot spots are no longer irrelevant at zero magnetic field, but reduce the zero-field lifetime by up to 5 times. We expect such a sizable effect to be observable as an upturn in $\rho_{ab}(T)$ while cooling through $T_{\mathrm{CDW}}$, in contrast to experiment.

The third model explores the main challenge set out at the start of this section; i.e. to maintain the RTA for cold charge, but let charge continue with a 50% probability after reaching a hotspot – to account for the non-collinear velocity – and continued cyclotron motion thereafter. The resulting quasiparticle trajectories form a cascade or waterfall as repeated encounters with hotspots fragment the distribution $P(k)$. The resulting MR shows a significantly increased turnover scale and reduced *H*-linear slope compared to the RTA model. Conceptually, these results are analogous with an effective increase in the distance between hotspots.

The fourth and final model combines small-angle scattering with the new hotspot scattering. This calculation preserves charge at all times and relies on the convergence of the distribution *P* to equilibrium such that at late times $\langle v\rangle$ vanishes. The result is a quadratic MR over the entire field range available to experiment. At fields above 30 T, LMR does persist, showing that ICM is still valid although the response is shifted to considerably higher magnetic field.

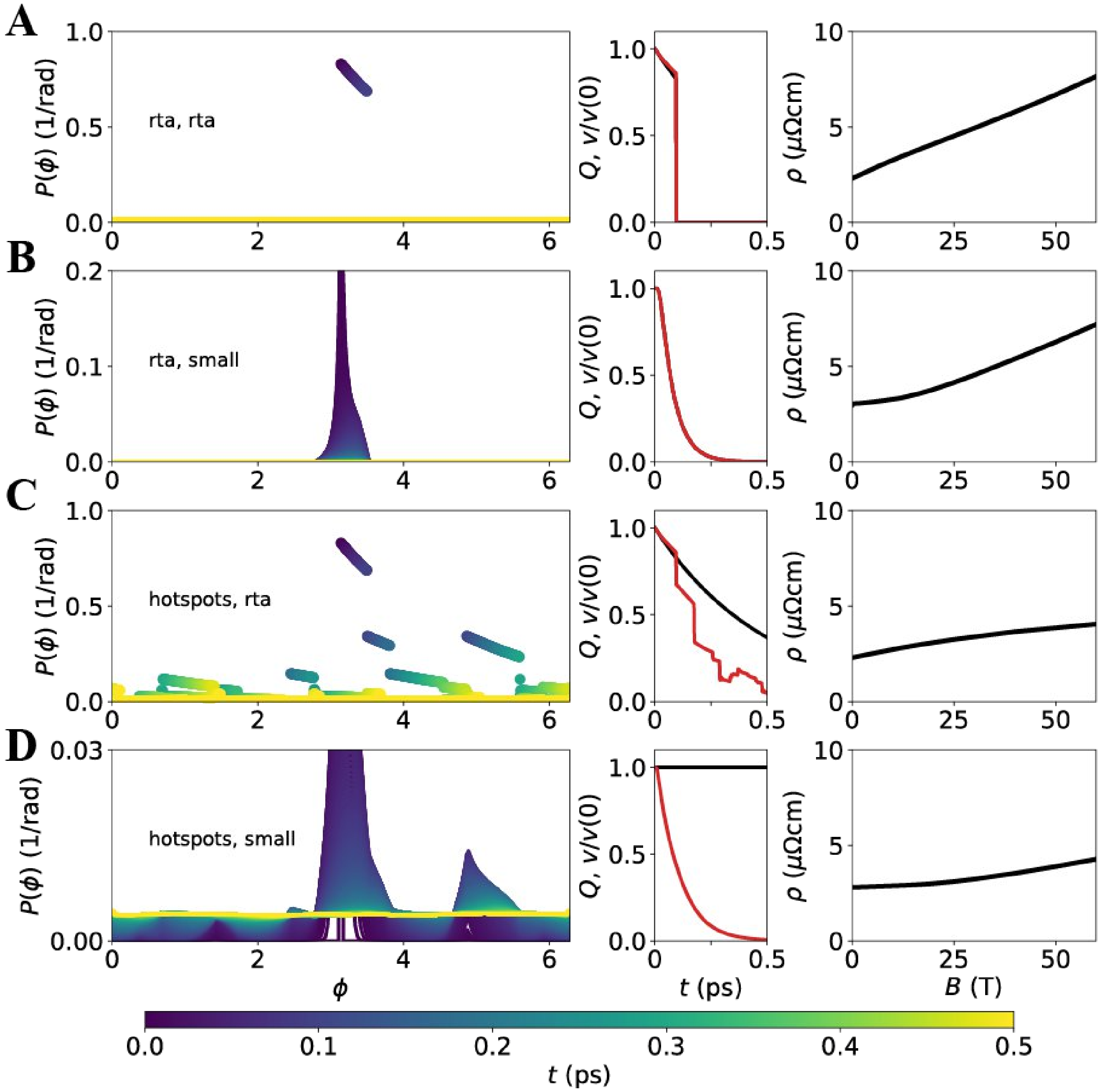

**Fig. S14**. Results beyond the RTA. A) Reproduction of the RTA results using the new formalism. The first panel shows the time evolution of $P(k)$ as a quasiparticle slowly decays until it reaches a hotspot and is impeded. The second column shows the total charge in the system (black) and the ensemble average velocity (red) over time. The third column shows the MR when the hot rate is sufficiently high to impede charge and the cold scattering rate matches the experimentally observed zero-field resistivity. B) Using small-angle scattering for the cold charge, the quasiparticle peak diffuses between nearby hotspots where charge terminates. The LMR persists but shows an extended quadratic regime at low magnetic fields. C) Using charge-conserving scattering between the ends of hotspots while keeping the RTA for cold scattering. The result is a complex cascade of repeated bifurcations in $P(k)$ whenever a hotspot is encountered. LMR remains, but with reduced slope and only above 25 T. D) Fully charge-conserving small-angle and hotspot scattering. LMR with reduced slope and increased turnover scale is obtained.

These four models, computed using the newly developed framework, show that the type of scattering for both the cold and hot charge has important consequences for the MR. Reminiscent of the work of Pippard on elemental copper, we find that the presence of sharp *k*-space features exaggerates the differences between small- and large-angle scattering, albeit here due to hotspots instead of sharp FS corners. Secondly, we find that small-angle scattering results in a longer quadratic regime at low magnetic fields while barely affecting the high-field LMR regime. The large-angle scattering assumed in the RTA is closer to the low turnover scale observed experimentally than small-angle scattering. Thirdly, we find that non-collinear velocities at (fluctuation-driven) hotspots maintain LMR, but with a strongly reduced slope and an increased turnover scale compared to the RTA version. Finally, combining small-angle and hotspot scattering we have demonstrated a fully charge-conserving model with LMR, albeit at a field strength that is too high, proving nonetheless that ICM does not violate charge conservation. Among these models, the RTA version which assumes large-angle cold scattering rate (likely due to point-like impurities in the planes) and a termination of charge at impedances (e.g. through spectral weight shifted away from the Fermi level) matches best with experiment. We emphasize more parameter space exists to be explored beyond the RTA, and the strong influence of details of the scattering processes keeps the possibility open to find a unified description for the MR and Hall effect in $NbSe_2$.

**Survey of other materials**

Given that the microscopic origin of ICM is not obvious, we compare here our results to other CDW materials and again find varying results. A number of compounds show MR compatible with a Peierls-like FS reconstruction. The sister compound 1*T*-$TiSe_2$ shows both QOs from the commensurate CDW reconstructed FS and a conventional MR (*70*,*71*). The commensurate CDW phase of 4*H*-$TaS_2$ also shows a saturating MR and QOs (*72*,*73*) and the same holds for 1*T*-$TaTe_2$ (*74*). The non-CDW compound 2*H*-$NbS_2$ shows a small quadratic MR with no signs of linearity, similar to what is observed in $NbSe_2$ outside the CDW phase (*75*). $NbSe_2$ is set apart among these dichalcogenides by its low $T_{CDW}$=33 K, the incommensurate nature of its CDW and the small fraction of density of states at the Fermi level affected by the CDW order.

In contrast to the above, a number of other compounds have shown behavior akin to our observations in $NbSe_2$. Thick films of $VSe_2$, for example, show a non-saturating *H*-linear MR with a slope of 0.11 $\mu\Omega$cm (of similar magnitude to that observed in $NbSe_2$) (*76*,*77*), though the disorder content in $VSe_2$ remains too high for any conclusion about QOs and possible saturation. In 2*H*-$TaS_2$ LMR was found which can be suppressed together with the commensurate CDW phase upon intercalation revealing QOs only outside the CDW phase despite the increase in disorder (*78*). A different report, however, shows QOs in the parent compound with unknown MR (*79*). The tritellurides $TbTe_3$ and $HoTe_3$ show LMR only inside the incommensurate CDW phase for which hotspots have been proposed as the origin (*80*,*81*), with studies extending to higher magnetic fields reporting magnetic breakdown and QOs (*82*). Finally, in $BaFe_2As_2$ not only the fluctuation regime near the QCP under doping, but also the spin density wave phase itself are also potential ICM candidates. Various groups have reported LMR which rapidly onsets just below $T_{SDW}$ (*3*,*5*,*83*), while another study using cleaner crystals has revealed a large, quadratic MR as well as QOs (*84*). Empirically, therefore, a role for finite disorder in combination with density wave order to realize LMR appears plausible, yet more systematic studies are required.

**Expectation of quantum oscillations in 2*H*-$NbSe_2$**

The real-space QO radius can be estimated via the semiclassical equation $l_0 = \hbar k_F/Bq$. Neglecting the less than 10% variation in FS radius $k_F$ across the Nb pockets, the cyclotron radius at 30 T is 75-120 nm for the six large, unreconstructed, Nb-derived pockets (see Table S1). In a FS reconstruction scenario for the CDW order, myriad QOs are expected with frequencies in the range 50-2000 Tesla, similar to what has been observed from 1 T in 4*H*-$TaS_2$, from 3 T in 2*H*-$TaS_2$ or to below 10 T in 2*H*-$TaSe_2$ (*72*,*79*). Such orbits have even smaller real-space radii down to 10 nm.

Comparing this characteristic scale to other measurements, we deduce that QOs are expected to be observed. Scanning tunneling microscopy studies have detected domains well above this size in clean $NbSe_2$ (*50*). The zero-field resistivity in our crystals is well described by a lifetime of 0.3 ps, which corresponds to a mean-free-path of 80 nm. While the mean-free-path relevant to QOs may, in principle, differ from this estimate, we remark here that the lifetime extracted from a Dingle analysis of the pancake pocket in the literature is in close agreement (*32*) and that $NbSe_2$ is neither strongly correlated nor a Dirac semimetal (scenarios where large differentials between the mean-free-path relevant to resistivity and QOs have been observed).

It is also instructive to compare with the observed QOs of the pancake pocket. Our measurements in Fig. S5 are to the best of our knowledge the first report of Shubnikov-de Haas oscillations in $NbSe_2$, i.e. measured in the resistivity, and are in good agreement with previous magnetization (de Haas-van Alphen) reports (*31*,*32*). Naturally, we cannot resolve QOs below the superconducting transition and strong oscillations emerge directly above $H_{c2}$ where the radius is reduced to approximately 50 nm (though the shape of these orbits is highly anisotropic). In the past, de Haas-van Alphen measurements have observed QOs below $H_{c2}$ with a large characteristic real-space radius of 250 nm starting at only 2 T (*31*,*32*). Under these conditions, the cyclotron orbits must be formed despite the presence of superconducting vortices. Indeed, experimentally the Dingle temperature was found to be 40% higher in the mixed state than beyond (*31*), suggesting even larger orbits could be observed in the absence of superconductivity.

All these estimates and comparisons indicate that QOs from the reconstructed FS are expected to be observable, yet none have been reported and we also observe none here up to 30 T. It will be intriguing to explore whether QOs emerge finally once $NbSe_2$ is tuned beyond the CDW phase, e.g. through the application of pressure.